\documentclass[
amsmath,
amssymb,
aps,
reprint,
pre,
showkeys,
onecolumn
]{revtex4-1}
\usepackage[utf8]{inputenc}
\usepackage{amsthm}
\usepackage{multirow}
\usepackage{hyperref}
\usepackage{graphicx}
\usepackage[font=small,justification=centerlast,format=plain]{caption}
\usepackage{subcaption}
\usepackage{xcolor}
\usepackage{dsfont}
\usepackage{booktabs}
\usepackage{bm}
\usepackage[capitalise]{cleveref}
\usepackage{comment}

\usepackage{mathrsfs}
\newcommand{\Hy} {\mathscr H}

\newcommand{\Lap}{\mathscr L}
\newcommand{\R}{\mathbb R}
\usepackage{array}

\begin{document}
	
	\title{Modelling Non-Linear Consensus Dynamics on Hypergraphs}
	\author{Rohit Sahasrabuddhe }
	\affiliation{Indian Institute of Science Education and Research, Pune, India}
	\author{Leonie Neuh{\"a}user}
	\affiliation{Hertie School Data Science Lab, Berlin, Germany}
	\author{Renaud Lambiotte}
	\affiliation{Mathematical Institute, University of Oxford, United Kingdom}
	\date{\today}
	
	\newtheorem{theorem}{Theorem}[section]
	\newtheorem{proposition}[theorem]{Proposition}
	\newtheorem{corollary}[theorem]{Corollary}
	\newtheorem{lemma}[theorem]{Lemma}
	\theoremstyle{definition}
	\newtheorem{definition}[theorem]{Definition}
	\newtheorem{assumption}{Assumption}
	\theoremstyle{remark}
	\newtheorem*{remark}{Remark}
	
	\renewcommand\theparagraph{(\alph{paragraph})}

\begin{abstract}

The basic interaction unit of many dynamical systems involves more than two nodes. In such situations where networks are not an appropriate modelling framework, it has recently become increasingly popular to turn to higher-order models, including hypergraphs. In this paper, we explore the non-linear dynamics of consensus on hypergraphs, allowing for interactions within hyperedges of any cardinality. After discussing the different ways in which non-linearities can be incorporated in the dynamical model, building on different sociological theories, we explore its mathematical properties and perform simulations to investigate them numerically. After focussing on synthetic hypergraphs, namely on block hypergraphs, we investigate the dynamics on real-world structures, and explore in detail the role of involvement and stubbornness on polarisation.
\end{abstract}

\maketitle	
	\section{Introduction} \label{sec:intro}
	
	Networks provide a powerful language to model systems made of interacting elements, as observed in many scientific domains. Within the social sciences, for instance, their significance has increased in recent years with the emergence of online social networks. In particular, social network analysis has become a common tool to help model and extract information from the myriad of social interactions in a system. In addition to algorithms extracting central nodes or clusters of similar nodes, much research has focussed on the impact of the network structure on the dynamics of opinion formation \cite{social_dyn_review}, looking at a variety of models, from linear ones, like the Voter model or the DeGroot model \cite{DeGroot1974}, to non-linear ones, like the  bounded-confidence models \cite{deffuant2000mixing} and threshold models \cite{watts2002simple}.
	
	Nevertheless, it is important to remember that networks are an abstract representation of real-world systems, and several studies have shown, in recent years, that they may  be inadequate to capture critical aspects of interacting systems \cite{lambiotte_2019,battiston2020networks}. In particular, the basic interacting units of a network are pairs of nodes, an assumption that is not verified in situations when multi-body or  group interactions take place, such as in neural activity \cite{giusti_clique_2015,reimann_cliques_2017, santos_topological_2018}, robotics \cite{olfati2007consensus} or scientific collaborations \cite{patania_shape_2017}. The importance of irreducible group interactions, that is interactions that can not be built from a combination of pairwise interactions, is especially relevant for opinion dynamics. Experiments in social psychology such as the conformity experiment \cite{asch_effects_1951} indicate that multiple exposures might be necessary for an agent to adopt a certain opinion state. This dependence on multiple contacts can be seen as a first step towards group effect. Threshold models, in which the state of agents switches if a certain fraction of their neighbours agree, have been developed to describe this phenomenon and are already inherently different from simple models of epidemic spread. However, these models are based on independent, pairwise interactions that are linearly accumulated and therefore do not account for truly higher-order effects \cite{chang_co_diff_2018}.	
	
	Popular choices for modelling genuine group interactions include hypergraphs \cite{random_hypergraphs_2019,berge_hypergraphs_1989,estrada_subgraph_2005} and simplicial complexes. The latter has opened the doors to the use of algebraic topology in the analysis and study of complex systems \cite{lambiotte_tda_2018, schaub_random_2018,mukherjee_random_2016,parzanchevski_simplicial_2017,muhammad_control_nodate,petri_simplicial_2018, salnikov2018simplicial}. 
	With regard to dynamical processes on higher-order models, several works have attempted to apply social dynamics to either of these structures \cite{iacopini_2019, petri_simplicial_2018,arruda2019social}.  Moreover, linear dynamics such as random walks and general spectral analysis have been extended to hypergraphs \cite{carletti_random_2020,helali_hitting_2019, zhou_learning_2006, lu_high-ordered_2011}. 
	Nevertheless, Neuhäuser et al. have shown with the introduction and analysis of the so-called three-body consensus model (3CM) \cite{neuhauser_multibody_2020} that multi-body dynamical effects that go beyond rescaled pairwise interactions can only appear if the interaction function is non-linear, regardless of the underlying multi-body structure. If the interaction function is linear, the system can always be written as a linear, pairwise interaction system on a rescaled network. Therefore, non-linearity is the essential ingredient to make genuine group dynamics appear.
	
	In this work, we generalise 3CM by proposing a Multi-body Consensus Model (MCM) for opinion consensus on hypergraphs of arbitrary group size, and explore different versions of interaction functions emphasising different sociological mechanisms, such as peer pressure and homophily, i.e. the tendency for similar nodes to interact more often. We then present the combined effects of these phenomena in numerical simulations on empirical hypergraphs. The rest of this paper is structured as follows. In Section \ref{sec:multi-body models}, we introduce the process of modelling multi-body interactions, and derive some useful analytical quantities. In Section \ref{sec:mcm}, we present and discuss MCM. Section \ref{sec:real-world} contains simulations on real-world hypergraphs and Section \ref{sec:discussion} is devoted to discussing further possibilities and avenues of research.
	\section{Modelling Multi-body Dynamical Systems}\label{sec:multi-body models}
			
	\subsection{Structure of a hypergraph}		
	The hypergraph $\Hy$ is a set $V(\Hy) = \{ 1, 2, \hdots, N \}$ of $N$ nodes, and a set $E(\Hy) = \{ E_1, E_2, \hdots, E_M \}$ of $M$ hyperedges. Each hyperedge $E_\alpha$ is a set of nodes, i.e. $E_\alpha \subseteq V(\Hy) \, \forall \, \alpha =1,2,\hdots,M$. We denote by $E^c(\Hy)$ the set of all hyperedges of cardinality $c$ (henceforth referred to as $c$-edges).
	
	We describe the structure of $\Hy$ using the adjacency tensors $A^c \in \R^{N ^ c}, \; c=2,3\hdots,N$, where $A^c$ represents the connections made by $c$-edges.
	\begin{equation}
	A^c_{ij\hdots} = \begin{cases}
	1 & \{ i, j\hdots \} \in E^c(\Hy)\\
	0 & otherwise
	\end{cases}
	\end{equation}	
	Thus $A^c$ is symmetric in all dimensions, and $A_{ij\hdots} = 1 \Rightarrow i \neq j \neq \hdots$.
	Each node $i \in V(\Hy)$ has a dynamical variable $x_i \in \R$ associated with it. $x_i$ is termed the 'state' of $i$, and represents the notion of an opinion of an individual $i$ in a hypergraph of social interactions.
	
	\subsection{General diffusion-like processes on hypergraphs}
	We denote by $\dot{x_i}^{E_\alpha}$ the effect of $E_\alpha$ on the derivative $\dot{x_i}$ of the state on node $i$. We have that $\dot{x_i}^{E_\alpha}=0$ when $i \not\in E_\alpha$, and for $i \in E_\alpha$, we write:
	\begin{equation}
	\dot{x_i}^{E_\alpha} = \sum_{j \in E_\alpha} s_{ijE_\alpha} \left( x_i, x_j, \hdots \right) (x_j - x_i)  \;\;\;\; i,j,\hdots \in E_\alpha
	\end{equation}
	This models the influence of $j\in E_\alpha$, $j \neq i$ on $i$ via the linear term $(x_j - x_i)$, modulated by the function $s_{ijE_\alpha}$. In this most general case, each triplet of affected node ($i$), acting node ($j$), and hyperedge ($E_\alpha$) is modulated by the unique function $s_{ijE_\alpha}$. Note that while the choice of modulating function is determined by this triplet, its argument is the states of all the nodes in $E_\alpha$. The complete 'diffusion-like' process is then obtained by linearly combining the effect of each hyperedge \cite{Srivastava2011,neuhauser_multibody_2020}, and one thus obtains the system of equations
	\begin{equation}
	\dot{x_i} = \sum_\alpha \dot{x_i}^{E_\alpha}.
	\end{equation}
	
	The special case where the choice of modulating function for a node $i$ is the same for all $E_\alpha$, and symmetric in all $k \in E_\alpha, k \neq i$ allows us to derive some interesting analytical results. Under this assumption, we denote the modulating function as $s_i$ and we can write the effect of all $c$-edges on $i$ as:
	\begin{equation}\label{eq:card_lap}
	\begin{split}
	\dot{x_i}^c &= \sum_{jk\hdots} A_{ijk\hdots}^c s_i(x_i, x_j, x_k, \hdots)\left( x_j - x_i + x_k - x_i + \hdots \right) \times \frac{1}{(c-1)!} \\
	&= \sum_{jk\hdots} A_{ijk\hdots}^c s_i(x_i, x_j, x_k, \hdots)\left(x_j - x_i\right) \times \frac{1}{(c-2)!}
	\end{split}
	\end{equation}
	Thus, $\dot{x_i}$ is given by:
	\begin{equation} \label{eq:master_lap} \dot{x_i} = \sum_{c=2}^N \sum_{jk\hdots} A_{ijk\hdots}^c s_i(x_i, x_j, x_k, \hdots)\left(x_j - x_i\right) \times \frac{1}{(c-2)!}
	\end{equation}
	
	\subsection{Deriving the Laplacian}\label{subsec:laplacian}
	We now derive the Laplacian for this process. Along the lines of \cite{neuhauser_multibody_2020}, we define weight matrices $W^c$ and the degree matrix $D^c$ as:
	\begin{equation}
	\begin{split}
	W_{ij}^c &= \sum_{kl\hdots} A_{ijk\hdots}^c s_i(x_i, x_j, \hdots) \frac{1}{(c-2)!} \\
	D_{ii}^c &= \sum_{jkl\hdots} A_{ijk\hdots}^c s_i(x_i, x_j, \hdots)\frac{1}{(c-1)!} = \sum_j \frac{W_{ij}^c }{(c-1)}
	\end{split}
	\end{equation}
	Here, $D_{ij}^c = 0 \; \forall\, i \neq j$. This allows us to write 
	\begin{equation}
	\dot{x_i}^c = - \sum_j \Lap_{ij}^c x_j
	\end{equation}
	where $\Lap_{ij}^c = (c-1)D_{ij}^c - W_{ij}$. Eq.(\ref{eq:master_lap}) can now be written as:
	\begin{equation}
	\dot{x_i} = - \sum_{c=2}^N \sum_j \Lap_{ij}^c x_j = - \sum_j \Lap_{ij} x_j 
	\end{equation}
	where $\Lap_{ij} = \sum_{c=2}^N \Lap_{ij}^c$.
	
	We see that when the modulation function $s_i$ is a constant, i.e. when the interactions are linear, the dynamics reduce to those of a static weighted network. However, when the interactions are non-linear, the corresponding network is time-dependent. Thus, we conclude as in \cite{neuhauser_multibody_2020} that irreducible multi-body effects are created only by non-linear interactions.
	
	\section{Multi-body Consensus Model}\label{sec:mcm}
	\subsection{Definition of the model}
	Following the discussions of the previous section, we introduce a general form of Multi-body Consensus Model (MCM) as follows:
	\begin{equation}
	\begin{split}
	\dot{x_i}^{E_\alpha} &= s^I_i \left( \mid \frac{\sum_{j \in E_\alpha} x_j}{\mid E_\alpha \mid} - x_i \mid \right) \\ 
	&\times \sum_{j \in E_\alpha} s^{II}_i\left( \mid \frac{\sum_{k \in E_\alpha, k \neq i} x_k}{|E_\alpha| - 1} - x_j \mid \right)(x_j - x_i) 
	\end{split}
	\end{equation}
	The effect of $E_\alpha$ on $i \in E_\alpha$ is modulated by two functions - $s^I_i$ and $s^{II}_i$. They define the two 'facets' of MCM, which can be studied separately in two simplified models that we call MCM I, with
	\begin{equation}\label{eq:mcm1}
	\dot{x_i}^{E_\alpha} = s_i ^ I \left( \mid \frac{\sum_{j \in E_\alpha} x_j}{\mid E_\alpha \mid} - x_i \mid \right) \left( \sum_{j \in E_\alpha} (x_j - x_i) \right),
	\end{equation}
	and MCM II, with
	\begin{equation}\label{eq:mcm2}
	\dot{x_i}^{E_\alpha} = \sum_{j \in E_\alpha} s^{II}_i\left( \mid \frac{\sum_{k \in E_\alpha, k \neq i} x_k}{|E_\alpha| - 1} - x_j \mid \right)(x_j - x_i).
	\end{equation}
	The difference in their sociological motivation and mathematical properties is captured by the arguments of their modulating functions.
	\begin{itemize}
		\item $s^I_i$ is a function of the distance of $x_i$ to the mean state of the hyperedge, and determines the rate at which that hyperedge influences its state.
		\item $s^{II}_i$ is a function of the distance of a participating node $j$ from the mean state of the hyperedge excluding $i$.
	\end{itemize}
	Thus, while MCM I modulates the competing effect of different hyperedges on the  state an incident node, MCM II determines which nodes inside a single hyperedge are the most influential. It is important to emphasise that $s^{II}_i$ reduces to the modulating function of the original 3CM for hyperedges of cardinality 3, whereas $s^I_i$ adds a new aspect that is not present in 3CM.
	In the next two sections, we discuss the mathematical and sociological differences of these functions further.

	\subsection{Mathematical Modelling Differences}
	
	\subsubsection{MCM I: Analysis in the mean field}\label{subsubsec:mcm1_analysis}	
	
	We note that both $s^I_i$ and $s^{II}_i$ are invariant under translations ($x_i \mapsto x_i + a$ for $a \in \R$) and reflection about the origin ($x_i \mapsto - x_i$). Thus, the dynamics are independent of the frame of reference in $\R$. Since the argument of $s^{II}_i$ is dependent on both $i$ and $j$, it cannot be analysed by the methods in Section \ref{subsec:laplacian}. In the case of $s^I_i$, in contrast, as it is symmetric in all $k \in E_\alpha, k \neq i$, its Laplacian can be derived as in Section \ref{subsec:laplacian}. Further, we can analytically derive some properties in the mean field.
	
	We assume that all nodes have the same modulating function, and denote it as $s^I$. Consider a hypergraph $\Hy$ with $m_k$ hyperedges of cardinality $k$ for $k=2,3,\hdots,N$. We assume the homogeneous mixing hypothesis, where the probability of a node participating in a hyperedge is independent of the node in question. Under this hypothesis, we can say that each node participates in $\frac{k m_k}{N}$ hyperedges of cardinality $k$ and that the mean of every hyperedge is the global mean $\bar{x}$. In order to examine the time evolution of $\bar{x}$ in the mean field, we proceed as follows.
	For $i \in E_\alpha$, we have
	\begin{equation}
	\dot{x_i}^{E_{\alpha}} = s^I(|\bar{x} - x_i|) \times |E_\alpha| (\bar{x} - x_i)
	\end{equation}
	Using this, we can write
	\begin{equation}
	\dot{x_i} = \sum_{k=2}^{N} \frac{k^2 m_k}{N} s^I(|\bar{x} - x_i|) (\bar{x} - x_i)
	\end{equation}
	Thus, the mean opinion evolves as
	\begin{equation}\label{eq:mean evolution}
	\dot{\bar{x}} = \frac{1}{N^2} \left( \sum_{k=2}^N k^2 m_k \right) \left( \sum_{i=1}^N s^I(|\bar{x} - x_i|) (\bar{x}-x_i) \right)
	\end{equation}
	and we observe that in a homogeneously mixed system, the mean does not shift if the distribution of $x_i$ about the mean is symmetric.
	
	We can also investigate the effect of asymmetry in the initial distribution of the states. Consider a situation where the initial states are binary (either 1 or 0). Suppose at $t=0$, $f_0$ fraction of the nodes have state $0$, and the rest ($f_1 = 1 - f_0$) have state $1$. From Eq.\ref{eq:mean evolution}, we can write
	\begin{align*}
		\dot{\bar{x}} &= \frac{1}{N^2}\left( \sum_{k=2}^N k^2 m_k \right) \left( \sum_{i=1}^N s^I(|f_1 - x_i|) (f_1 - x_i) \right)\\
		&= \frac{1}{N}\left( \sum_{k=2}^N k^2 m_k \right) f_0 f_1 (s^I(f_1) - s^I(f_0))
	\end{align*}
	If $s^I$ is monotonically increasing, $f_1 > f_0$ implies that $\dot{\bar{x}} > 0$ and $f_1 < f_0$ that $\dot{\bar{x}} < 0$, i.e. $\bar{x}$ shifts towards the majority. Similarly, $\bar{x}$ shifts towards the minority for monotonically decreasing $s^I$.

	\subsubsection{Numerical Simulations}
	To demonstrate the fundamental differences between MCM I and II, we run numerical simulations of each model on identical topologies, and with the same choice of modulating function.
	As a natural choice for modulation that is monotonic, continuous, and always positive, we consider for this illustration exponential functions.
	\begin{equation}
	\begin{split}
	s^I_i (x) = e^{\lambda x} \;\; \forall i \in V(\Hy)\\
	s^{II}_i (x) = e^{\delta x} \;\; \forall i \in V(\Hy)
	\end{split}
	\end{equation}
	We define a hypergraph with $N$ nodes as fully connected if all $2^N-N-1$ hyperedges with cardinality $\geq 2$ exist. We simulate the models on a fully connected hypergraph with $N=10$ nodes by initialising node states as  binary numbers ($0$ or $1$),  with $n_0$ nodes of state $0$. Since the modulating functions are always positive, a global consensus is reached without oscillations as expected. In the simulations, we define the global consensus as the average state when the standard deviation $\sigma^2(\{x_i\}) \leq 0.0005$.
	\begin{figure}
		\centering
		\begin{subfigure}{0.45\linewidth}
			\includegraphics[width=\textwidth]{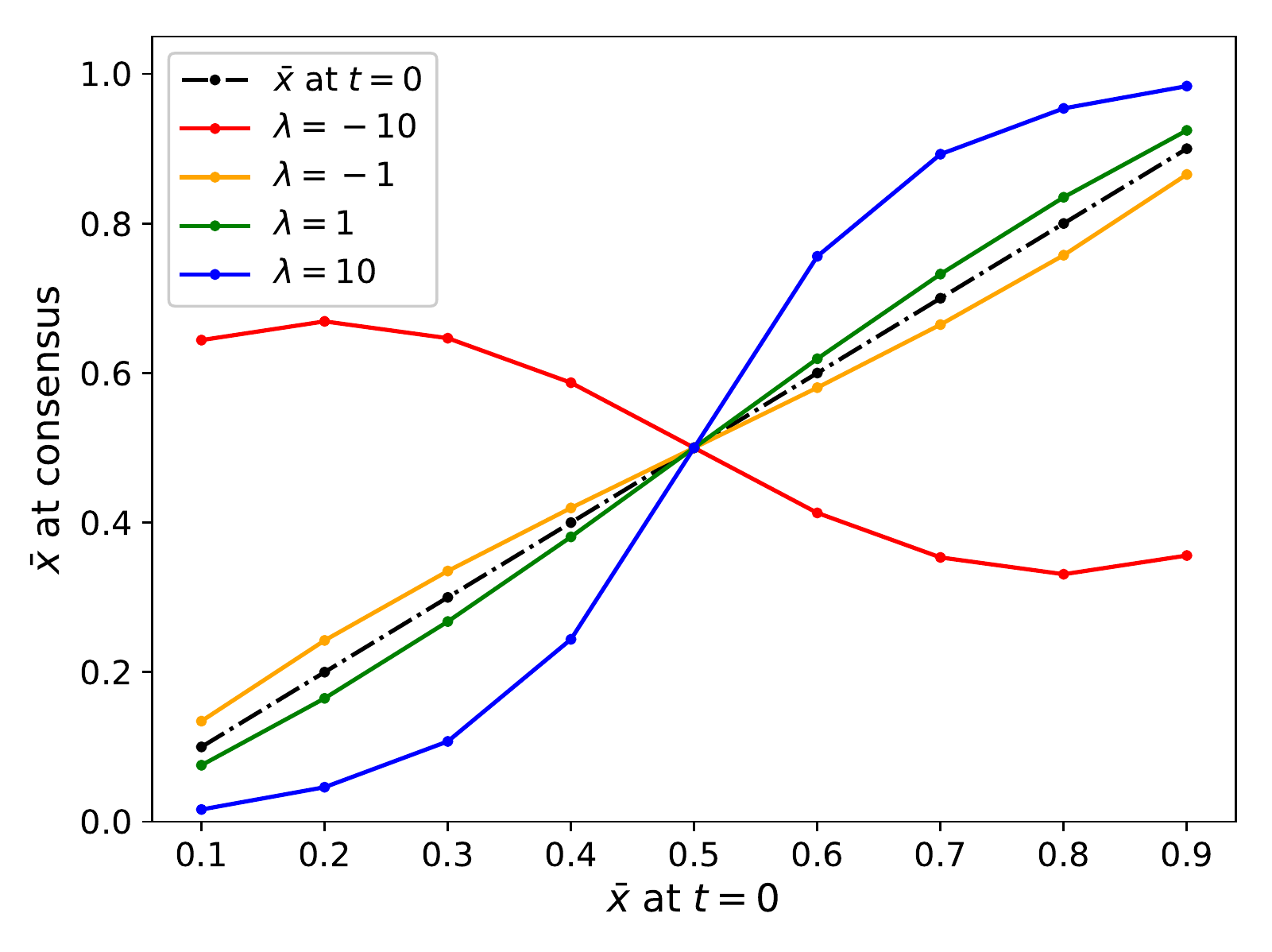}
		\end{subfigure}
		\begin{subfigure}{0.45\linewidth}
			\includegraphics[width=\textwidth]{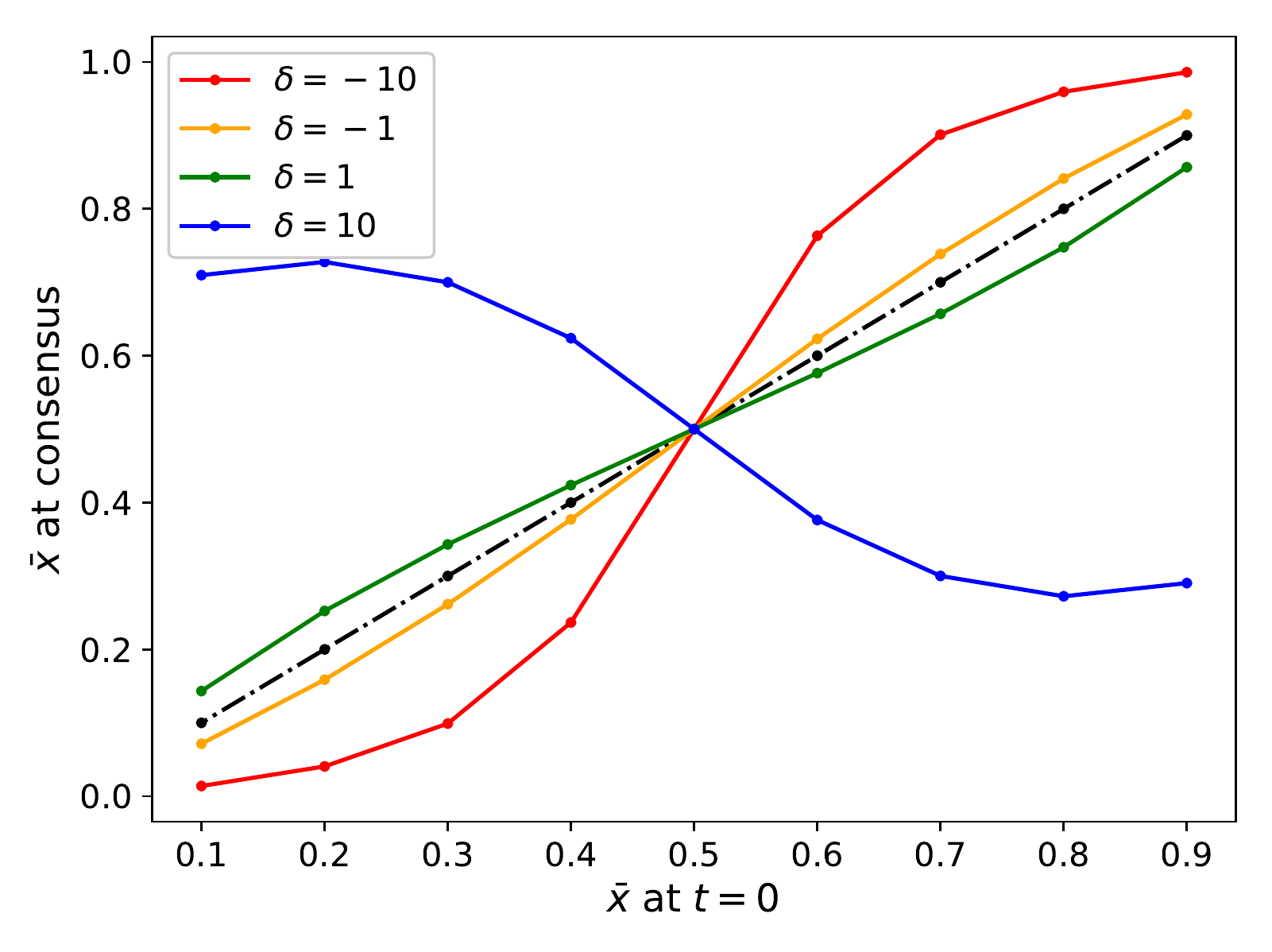}
		\end{subfigure}
		\caption{Numerical simulations to compare the evolution of MCM I (left) and II (right) on a fully connected hypergraph.}
		\label{fig:contrast}
	\end{figure}
	
	Numerical results in Fig. \ref{fig:contrast} show that the two facets evolve in completely contrasting ways. Further, the results for MCM I validate the analytical results in Section \ref{subsubsec:mcm1_analysis}. For a monotonically increasing modulating function ($\lambda > 0$), we see that the mean state shifts towards the initial majority. Similarly, it shifts towards the initial minority for a monotonically decreasing modulating function ($\lambda < 0$). While MCM II is a direct generalisation of 3CM, MCM I shows  an opposite behaviour, despite the same choice function for the modulating function. These drastic differences underline the distinct nature of the two modulating functions, -- and of their associated models MCM I and MCM II --, as well as the huge effect of the argument in the modulation function on the modelling outcome.  After this first analysis of the mathematical differences between the two variations of MCM, let us now turn to their sociological motivation, in order to chose appropriate modulating functions.
	
	\subsection{Sociological Differences}
	
	\subsubsection{MCM I}
\begin{figure}
		\centering
		\includegraphics[width=0.45\textwidth]{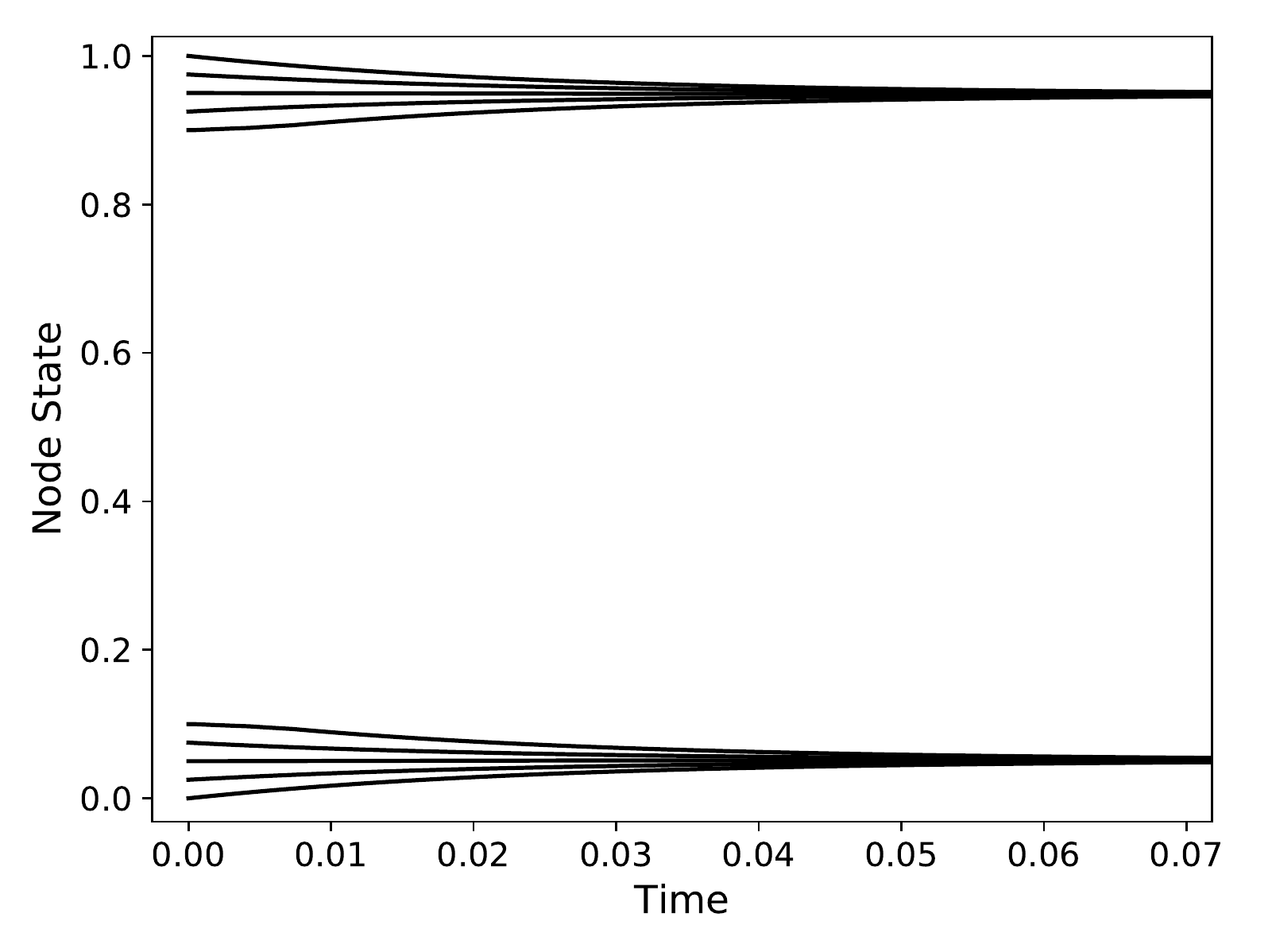}
		\caption{MCM I with the step modulating function ($\phi=0.1$) for a fully connected hypergraph with $N=10$}
		\label{fig:homo_mcm1}
	\end{figure}

	Homophily is a central concept in sociology  describing the tendency of like-minded individuals to interact \cite{homophily}. The topology of social interactions	is often influenced heavily by homophily. In sociological terms, the argument to $s^I_i$ quantifies the difference between the opinion of individual $i$ and the average opinion of group $E_\alpha$ that $i$ belongs to. The influence of a group on a node is thus determined by the proximity of its average state to the state of the node. For instance, consider the step function:
	\begin{equation}
	s^I_i(x) = \begin{cases}
	1 & x \leq \phi_i \\
	0 & otherwise 
	\end{cases}
	\end{equation}	
	Individual $i$ is influenced by a group only if their opinions differ by a value less than the threshold $\phi_i$. This mechanism is reminiscent of threshold models \cite{granovetter} \cite{watts} and bounded confidence models like the Deffuant model \cite{deffuant2000mixing}, with the important difference that it is group-specific, i.e. each hyperedge corresponds to one group, while network-based models usually consider the whole neighbourhood of a node as its single group.
	As an illustration, in Fig. \ref{fig:homo_mcm1}, we present the evolution of MCM I on a fully connected hypergraph ($N=10$) initialised as two tight groups around $0.05$ and $0.95$ with $\phi_i=0.1$ for all nodes. We see that no global consensus is reached despite the existence of groups where individuals of vastly different opinions may interact.
	
\begin{figure}
		\centering
		\begin{subfigure}{0.45\linewidth}
			\includegraphics[width=\textwidth]{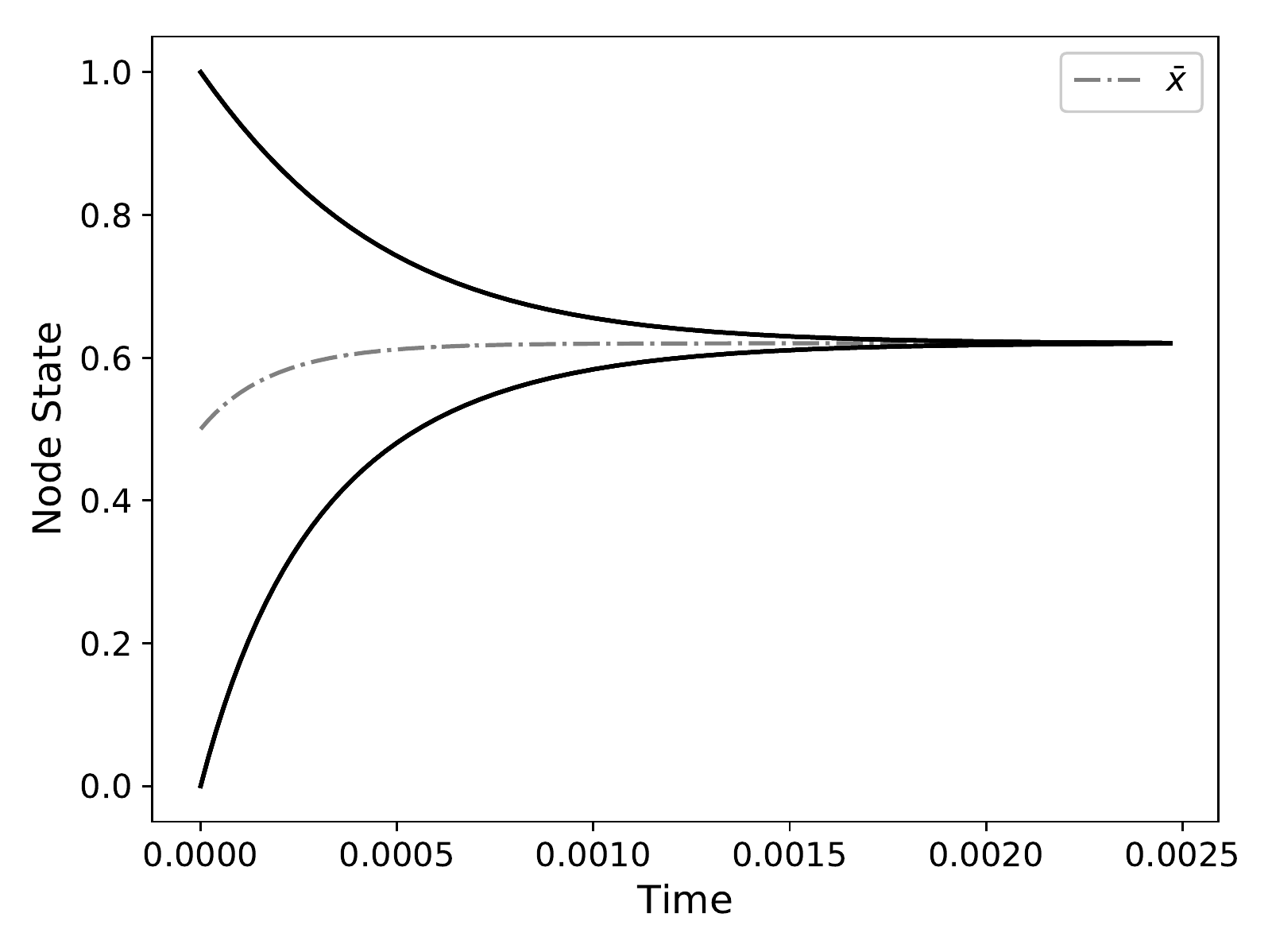}
			\subcaption{Evolution with $\Delta=1$}
		\end{subfigure}
		\begin{subfigure}{0.45\linewidth}
			\includegraphics[width=\textwidth]{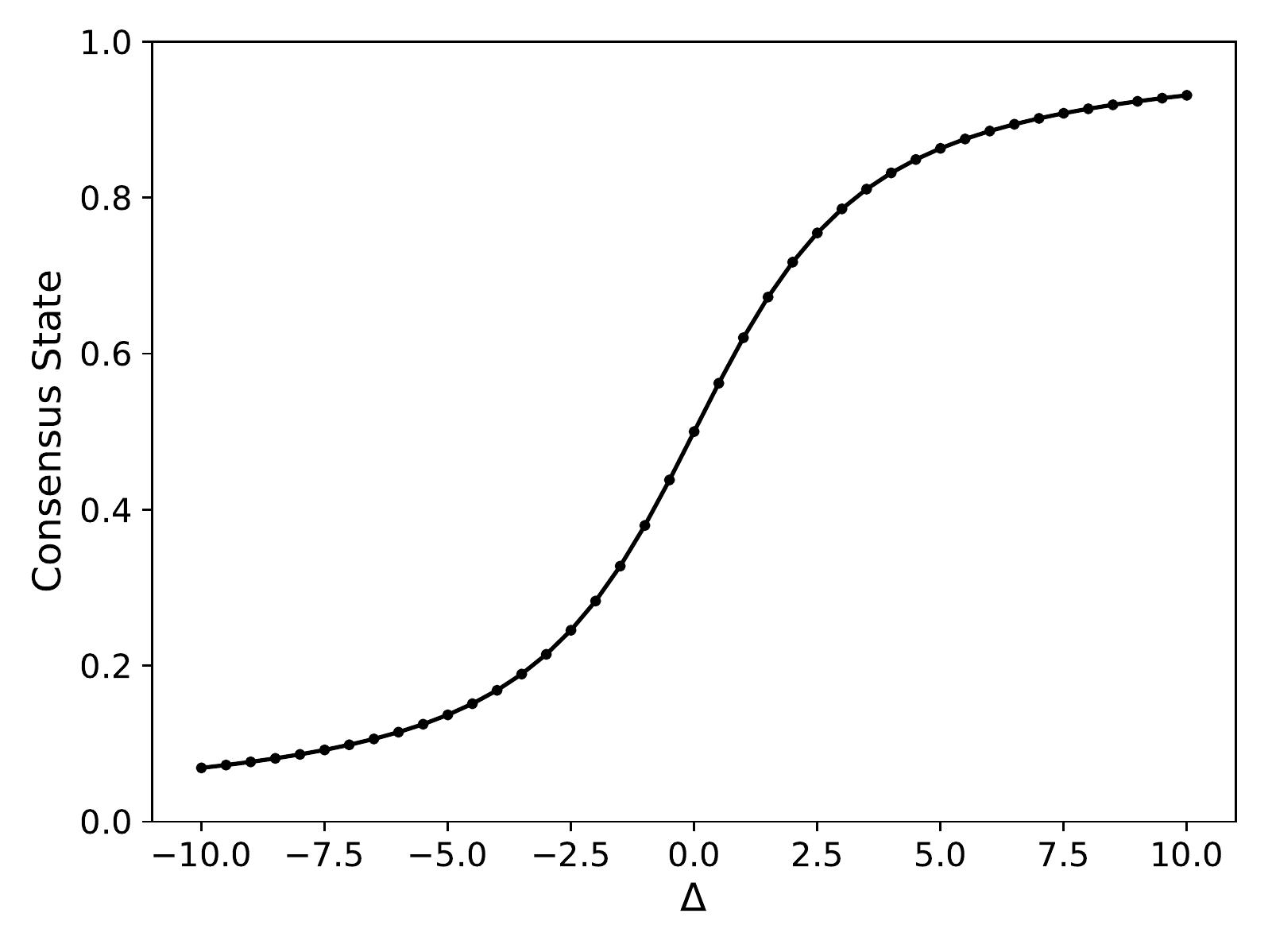}
			\subcaption{Consensus vs $\Delta$}
		\end{subfigure}
		\caption{Evolution of MCM I on a fully connected hypergraph of $10$ nodes initialised with $5$ nodes each of opinions $0$ (with $\lambda_i = \Delta$) and $1$ (with $\lambda_i = -\Delta$).}
		\label{fig:stubborn_cc}
	\end{figure}

	Another choice for $s^I_i$ is a monotonic function, where a decreasing function represents an individual $i$ who is less influenced by groups with opinions very different from its own than by groups with similar opinions. This can be thought of as individual $i$ resisting change, or some form of 'stubbornness'. On the other hand, an increasing function can be thought of as representing 'gullibility'. Consider the exponential function:
	\begin{equation}
	s^I_i(x) = e^{\lambda_i x}
	\end{equation}
	For $\lambda_i < 0$ ($>0$), the function is monotonically decreasing (increasing). Stubborn (gullible) nodes are therefore characterised by $\lambda_i < 0$ ($>0$).
	In Fig. \ref{fig:stubborn_cc}, we present the evolution of MCM I with an exponential modulation	on a fully connected hypergraph ($N=10$) with binary symmetric initialisation. The nodes initialised to $1$ ($0$) have $\lambda_i = -\Delta$ ($\Delta$). Numerical results show that consensus shifts towards the opinion of stubborn individuals.
	
	An important effect of MCM I is that it makes it possible for an individual to heavily influence other members of a group while being resistant to their influence. This allows certain individuals to be 'trendsetters' and to pull entire groups towards their opinion. Stubborn individuals in a group of people with whom they disagree can be trendsetters.
	
	\subsubsection{MCM II}\label{subsec:soc_mcm2}
	The pressure to conform is an important sociological and psychological phenomenon \cite{asch}, at the root of most models of opinion dynamics. 'Conformity' is used to describe the tendency of an individual to align its beliefs to those of its peers, and is usually affected by the reinforcing nature of shared opinions (peer pressure). The sociological interpretation of the argument of $s^{II}_i$ is the difference between the opinion of individual $j$ to the average opinion of the group except individual $i$. Thus, in MCM II,  the influence exerted by $j$ inside a hyperedge depends on the proximity of its opinion to those of the rest of the group. For instance, consider the exponential function:
	\begin{equation}
	s^{II}_i (x) = e^{\delta_i x}
	\end{equation}	
	When $\delta_i < 0$, individual $i$ tends to be more influenced by individuals who agree with the rest of the group. In contrast, for $\delta_i > 0$ an individual is attracted to the outliers of a group. 'Anti-conformists' or 'contrarians' are individuals whose decisions oppose the majority, and can thus be associated with $\delta_i > 0$.
	In Fig. \ref{fig:contrast} (b), we see that in a population of conformers (contrarians), the mean shifts towards the initial majority (minority), as expected.

	\subsection{Defining $s^I_i$ and $s^{II}_i$ for MCM}
	\label{sec:MCM1MCM2}
	The previous sections considered MCM I and MCM II separetedly. Let us now consider a general model integrating both types of mechanisms.
	To do so, we build on Social Judgement Theory \cite{sjt} where the responses of people to different opinions are categorised as:
	\begin{itemize}
		\item The latitude of acceptance - where the other opinion is sufficiently close to their own belief and is accepted. Their beliefs shift towards the new opinion (assimilation).
		\item The latitude of rejection - where the other opinion is too far from theirs and is therefore unacceptable. This leads to their opinion shifting away from the other (contrast).
		\item The latitude of non-commitment - where the other opinion is neither sufficiently close to be accepted nor different enough to be repulsive.
	\end{itemize}
	According to Social Judgement Theory, the more the individual is personally involved in an issue, the smaller their latitudes of acceptance and non-commitment.
	Inspired by this categorisation, and drawing from the Jager-Amblard model \cite{jager-amblard}, we define for each node $i$, two thresholds - that of acceptance ($\phi^A_i$) and that of rejection ($\phi^R_i$), and we propose
	\begin{equation}
	\label{oppo}
	s^I_i (x) = \begin{cases}
	e^{\lambda_i x} & x \leq \phi^A_i\\
	0 & \phi^A_i < x < \phi^R_i\\
	-e^{\lambda_i x} & x \geq \phi^R_i
	\end{cases}
	\end{equation}
	For MCM II, we keep an exponential modulating function.
	\begin{equation}
	s^{II}_i (x) = e^{\delta_i x}
	\end{equation}
	
	\subsection{Evolution of MCM on a block hypergraph}
\begin{figure}
		\includegraphics[width=0.45\textwidth]{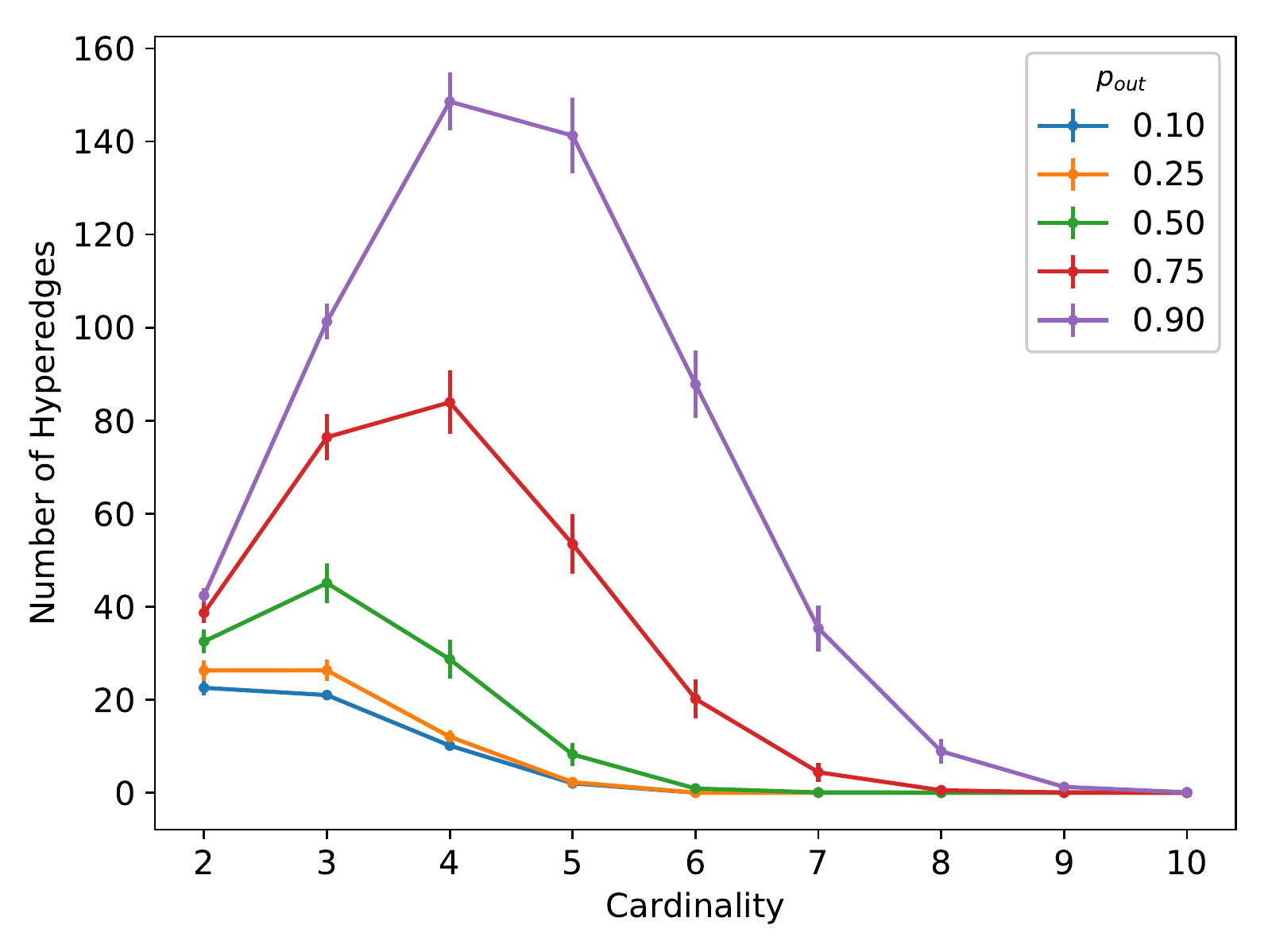}	
		\caption{Cardinality distribution for a block model with $N=5$ for various values of $p_{out}$.}
		\label{fig:card_dist_block}
	\end{figure}

	In the previous sections, simulations were run on fully connected hypergraphs. Let us now turn our attention to more nuanced hypergraph models in order to uncover the impact of structure on dynamics. As a first step, we consider a block hypergraph model, inspired by the stochastic block network model, and defined for a hypergraph with $2$ blocks of $N$ nodes each. 
	The probability of a hyperedge with $n_1$ and $n_2$ nodes from the two blocks respectively being created is given by $p_{out}^{n_1 n_2}$, where $p_{out}$ is a parameter for the model. Note that intra-block hyperedges have either $n_1=0$ or $n_2=0$, and are thus created with probability 1. This means that both blocks are fully connected, and the number of inter-block hyperedges can varied by tuning $p_{out}$.
	In Fig. \ref{fig:card_dist_block}, we plot the cardinality distribution of a block model of a hypergraph averaged over $1000$ instances with $N=5$.

	The system has several parameters, some associated with its structure, and other to the dynamical model.	
	As a first descriptive study, we study the effect of $\phi_i^A$ and $\phi_i^R$, we fix $\delta_i=-5$, $\lambda_i=-1$ for all nodes. We consider a block hypergraph with $N=5$ and $p_{out}=0.4$, with the initial opinions of the nodes in the two blocks as values from a uniform distribution over $[0, 0.5]$ and $[0.5, 1]$ respectively. Further we set $\phi_i^A = \phi^A$ and $\phi_i^R = \phi^R$ as the same for all nodes. For each choice of the parameters $\phi_i^A$ and $\phi_i^R$, we simulate $500$ realisations of the evolution of MCM. Unlike the numerical simulations performed previously, MCM is not guaranteed to reach a global consensus. Here, we stop the simulation when the difference in standard deviation over $500$ timesteps is less than $0.0005$. 
	
\begin{figure}
		\centering
		\begin{subfigure}{0.45\linewidth}
			\includegraphics[width=\textwidth]{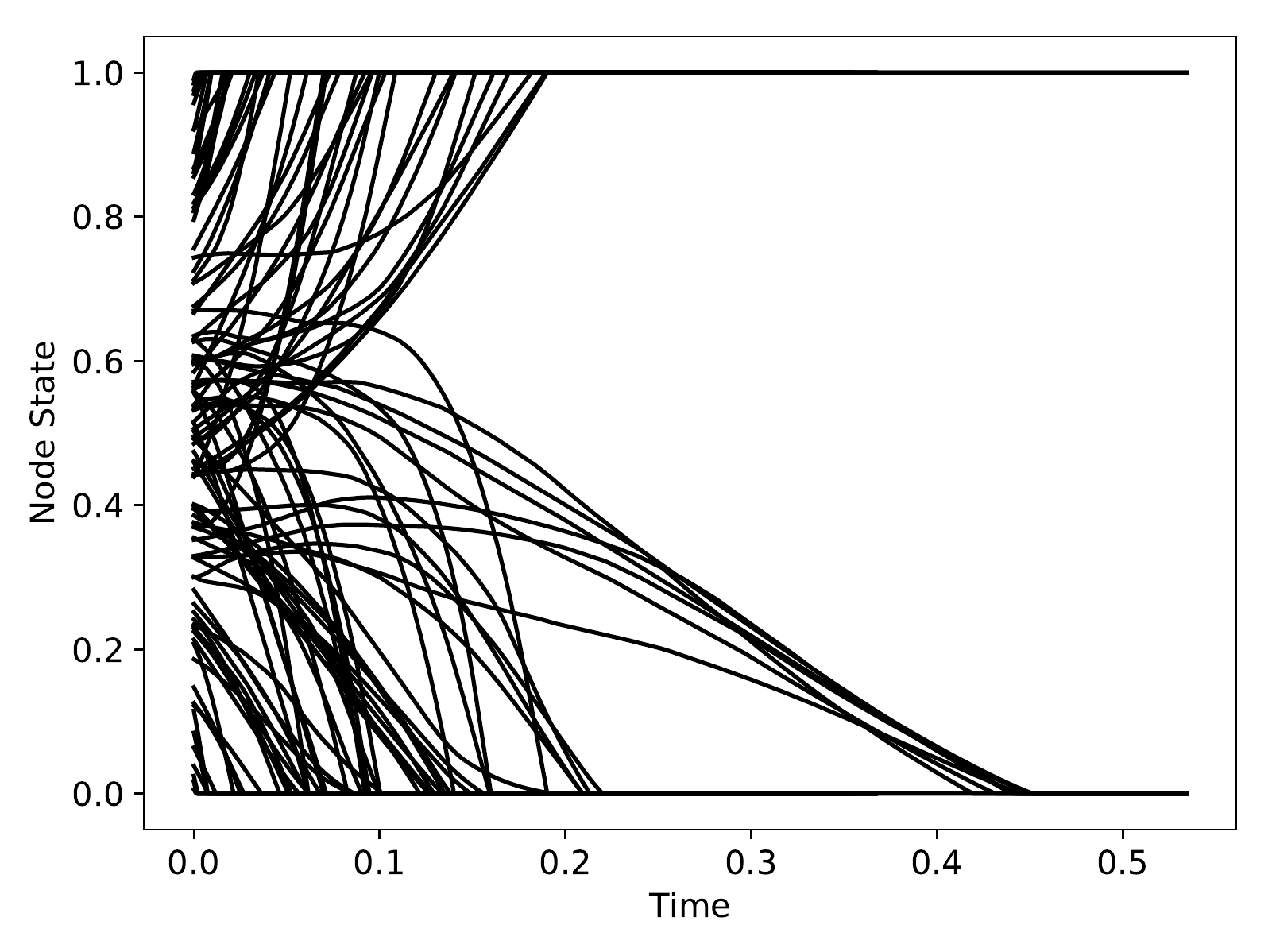}
			\subcaption{$\phi^A = 0.10, \phi^R = 0.15$}
		\end{subfigure}
		\begin{subfigure}{0.45\linewidth}
			\includegraphics[width=\textwidth]{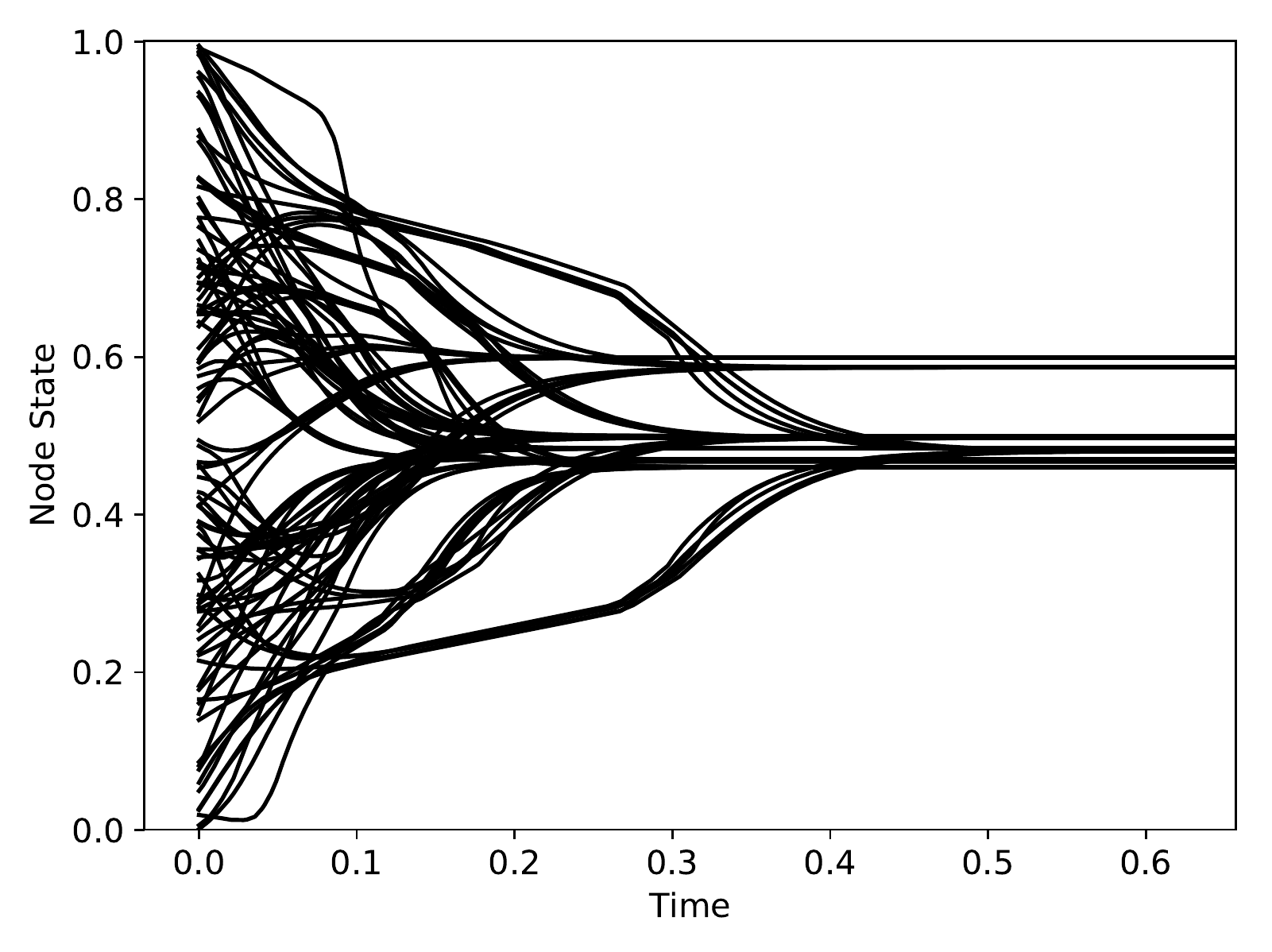}
			\subcaption{$\phi^A = 0.20, \phi^R = 0.60$}
		\end{subfigure}\\
		\begin{subfigure}{0.45\linewidth}
			\includegraphics[width=\textwidth]{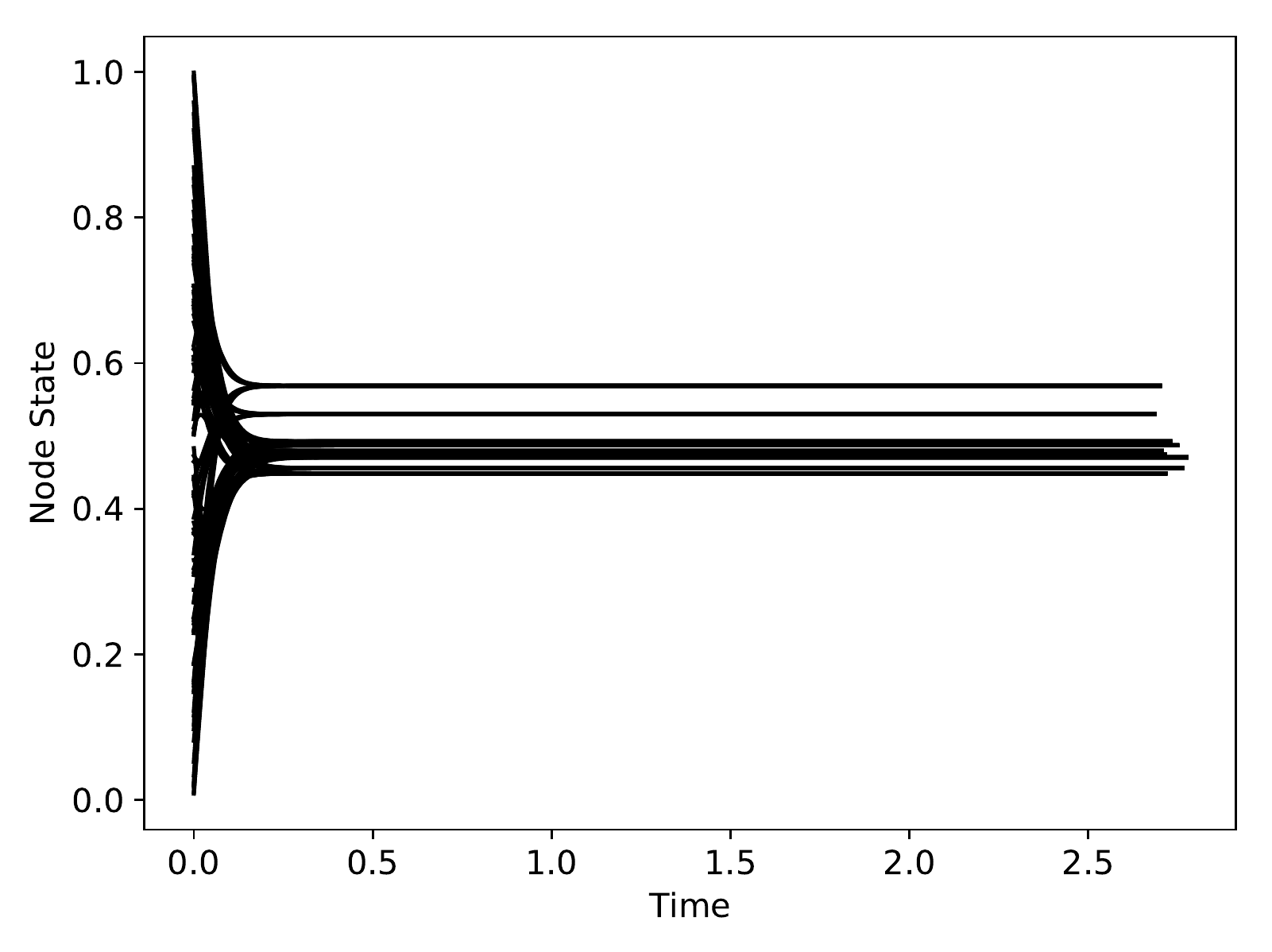}
			\subcaption{$\phi^A = 0.40, \phi^R = 0.80$}
		\end{subfigure}
		\caption{Typical evolution of MCM on the block hypergraph with $N=5$, $p_{out}=0.4$}
		
		\label{fig:block_polarisation}
	\end{figure}

	In Fig. \ref{fig:block_polarisation}, we plot the evolution of $10$ (out of the $500$) realisations for each set of values for the parameters. 
	This preliminary analysis reveals that low values of $\phi^R$, which represents intolerance in the social context, create polarisation. In contrast, a large value of $\phi^A$ encourages consensus. In all the simulations, the range of possible opinions is restricted to $[0,1]$. This allows us to define the notion of extreme nodes as those with state $> 0.9$ or $<0.1$, and thus to quantify the extent of polarisation. This interpretation is confirmed by the results in Table \ref{tab:block_polarisation}, where we see that for low (high) values of $\phi^A$ and $\phi^R$, there is polarisation (consensus). For intermediate values, the two blocks reach an internal consensus first, before converging to the global consensus. Due to the small size of the blocks, as well as the presence of several inter-block hyperedges, we do not see a co-existence of extreme and non-extreme nodes. However, for large group sizes and more sparsely connected blocks, this coexistence is indeed possible.

	\begin{table}
		\centering
		{\renewcommand{\arraystretch}{1.5}
			\begin{tabular}{|c|c|c|}
				\hline
				$\phi^A$ & $\phi^R$ & Fraction of extreme nodes\\
				\hline
				0.10 & 0.15 & $ 0.9990 \pm 0.0099$ \\
				0.20 & 0.60 & $0.0 \pm 0.0$ \\
				0.40 & 0.80 & $0.0 \pm 0.0$ \\
				\hline	
		\end{tabular}}
		\caption{Evolution of MCM on the block hypergraph with $N=5$, $p_{out}=0.4$}
		\label{tab:block_polarisation}
	\end{table}	
	
	In the next sections, we will explore the dynamics of MCM for these specific choices of interaction function, with a specific focus on the emergence of polarisation and the importance of stubbornness.
	
	\section{Numerical simulations on real-world hypergraphs}\label{sec:real-world}
	\subsection{Datasets}
	
\begin{figure}
		\begin{subfigure}{\linewidth}
			\includegraphics[width=0.4\textwidth]{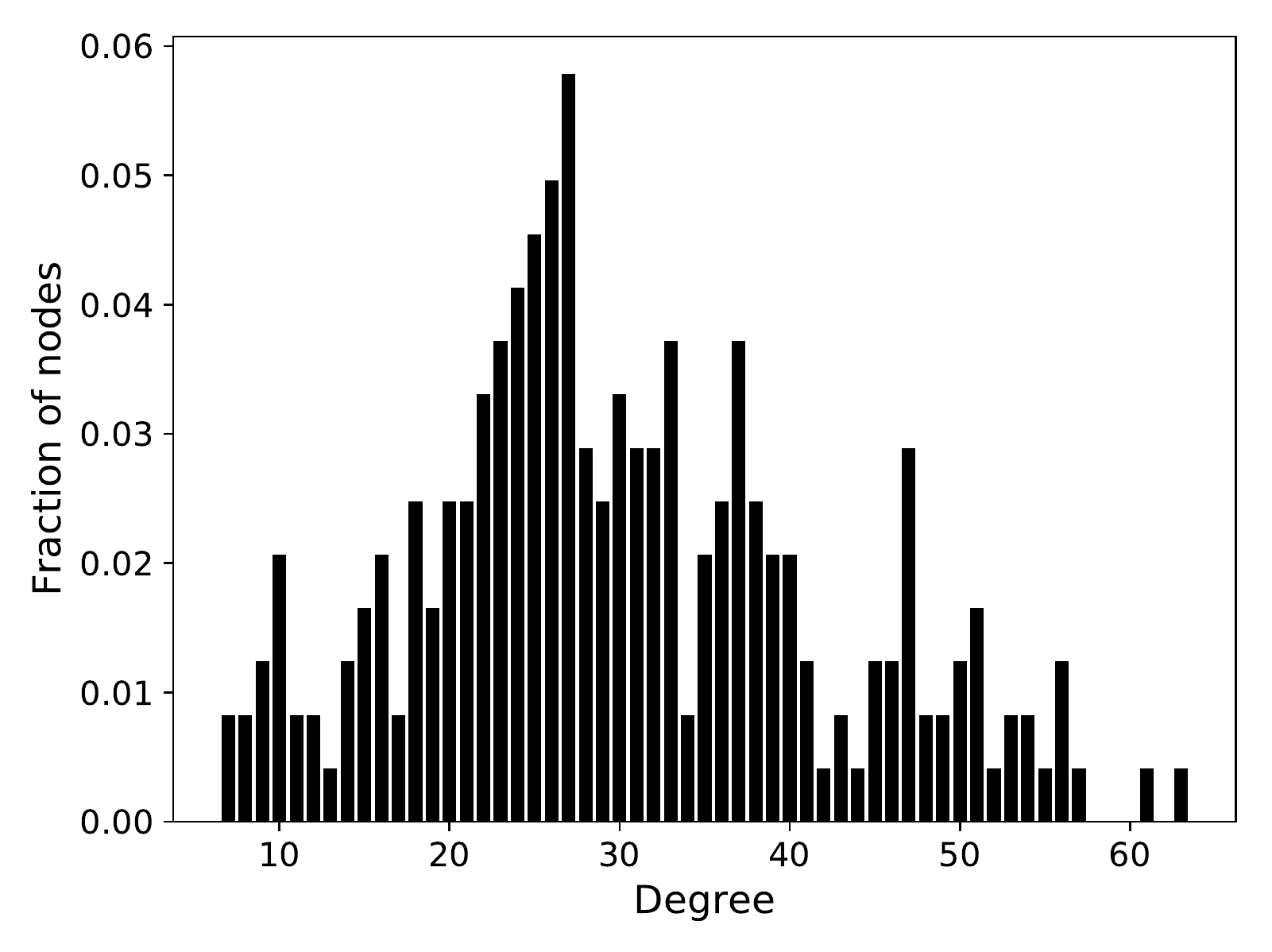}
			\includegraphics[width=0.4\textwidth]{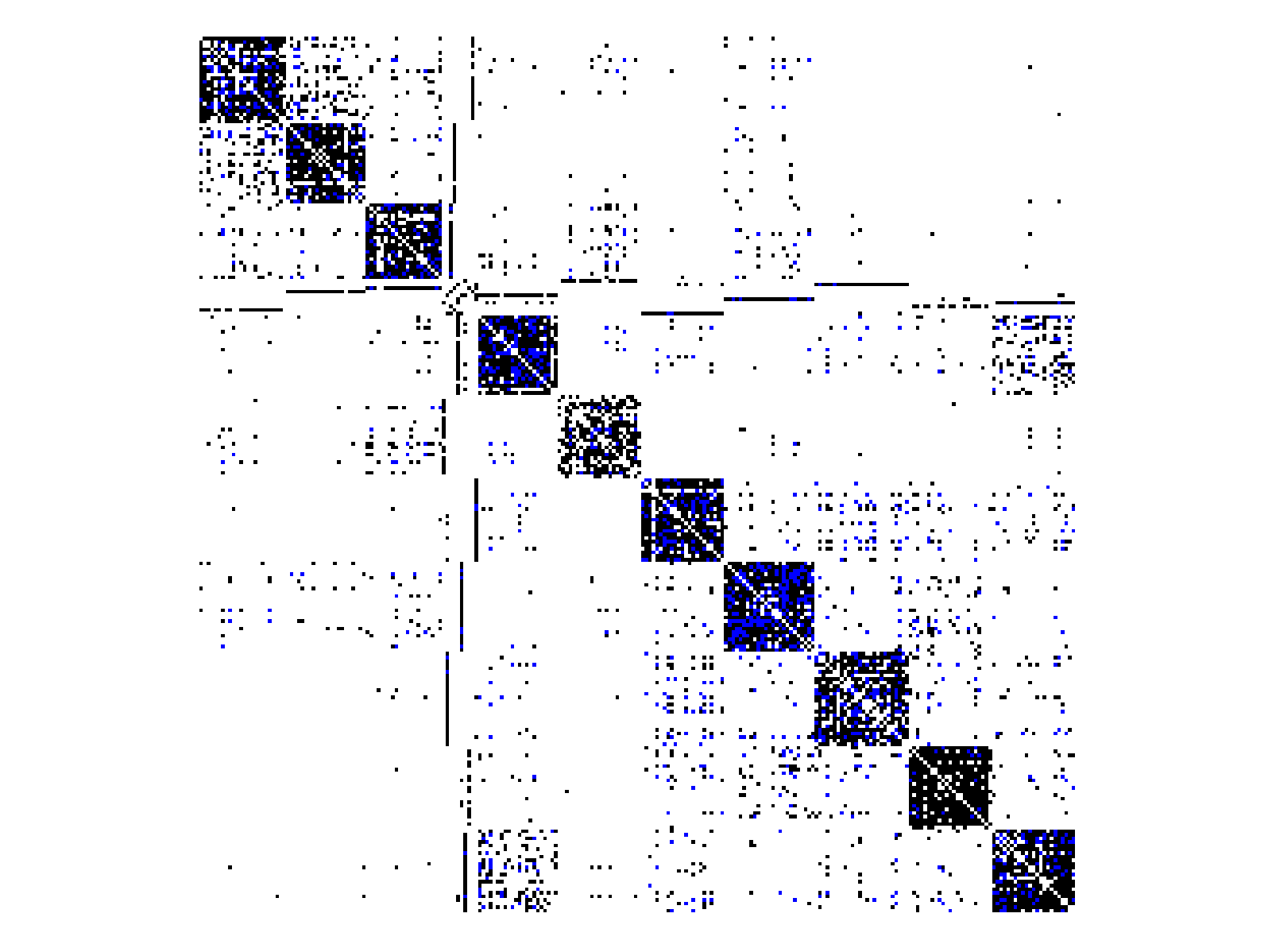}
			\subcaption{Primary School}
		\end{subfigure}\\
		\begin{subfigure}{\linewidth}
			\includegraphics[width=0.4\textwidth]{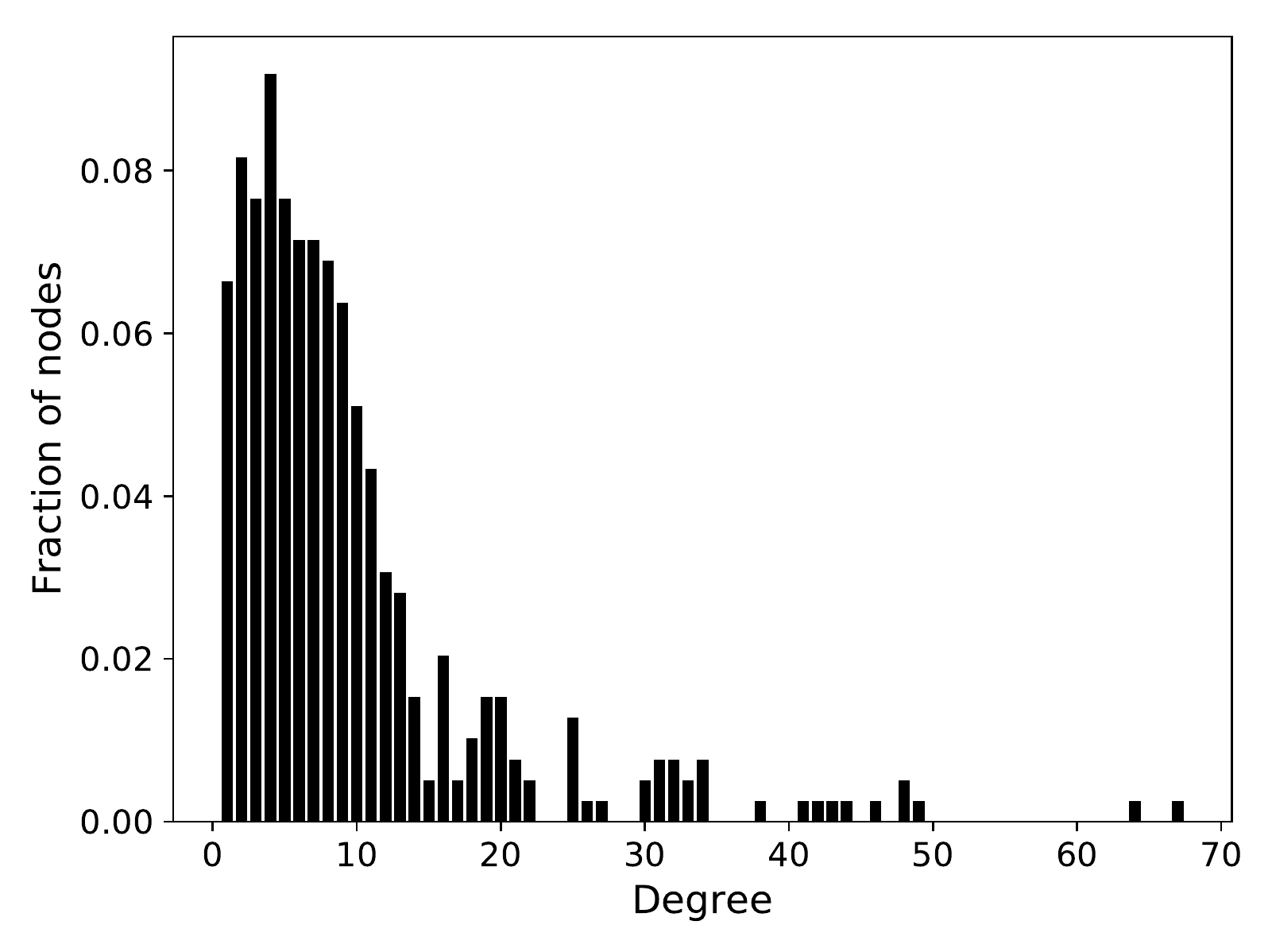}
			\includegraphics[width=0.4\textwidth]{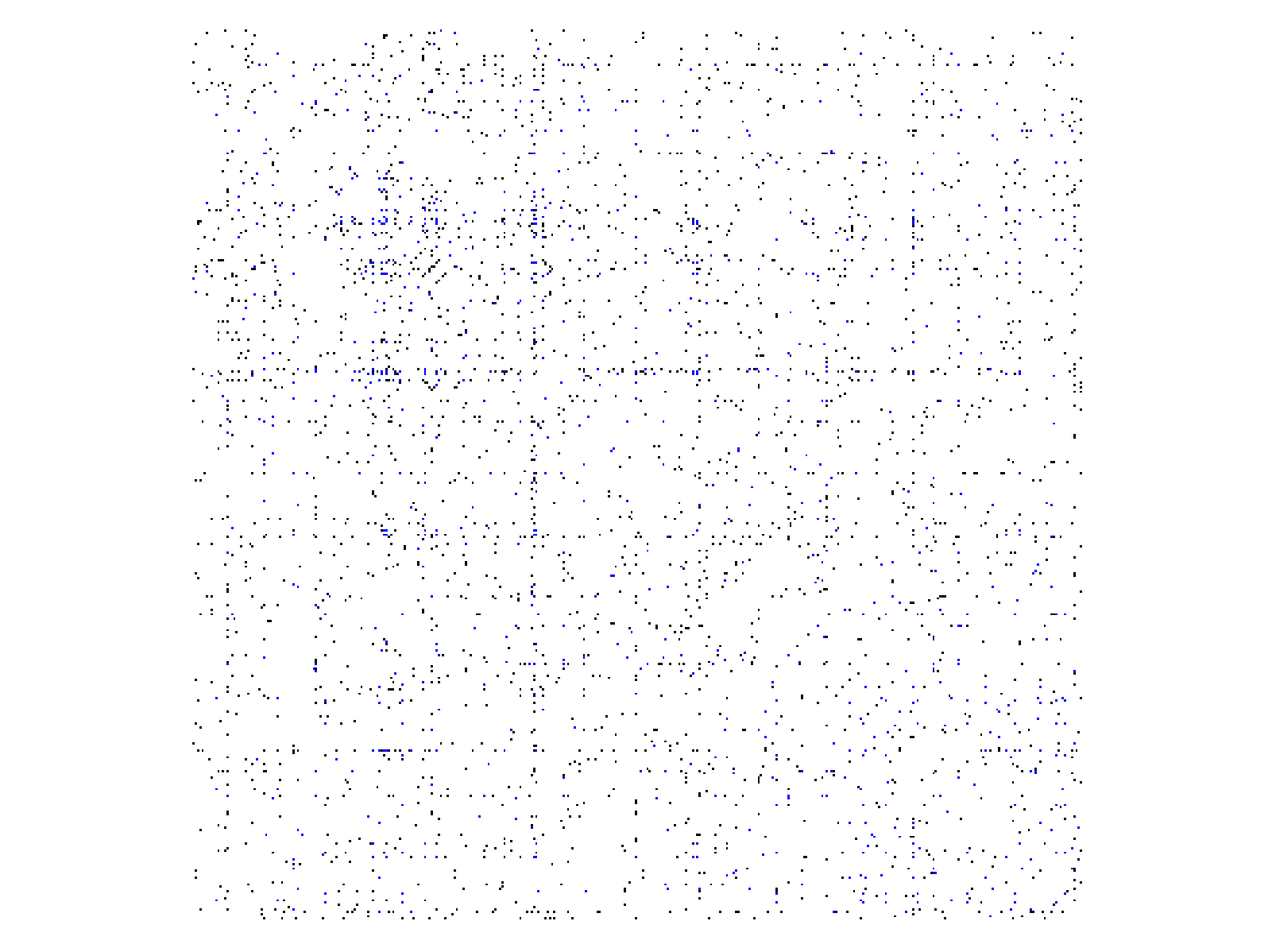}
			\subcaption{Conference}
		\end{subfigure}\\
		\begin{subfigure}{\linewidth}
			\includegraphics[width=0.4\textwidth]{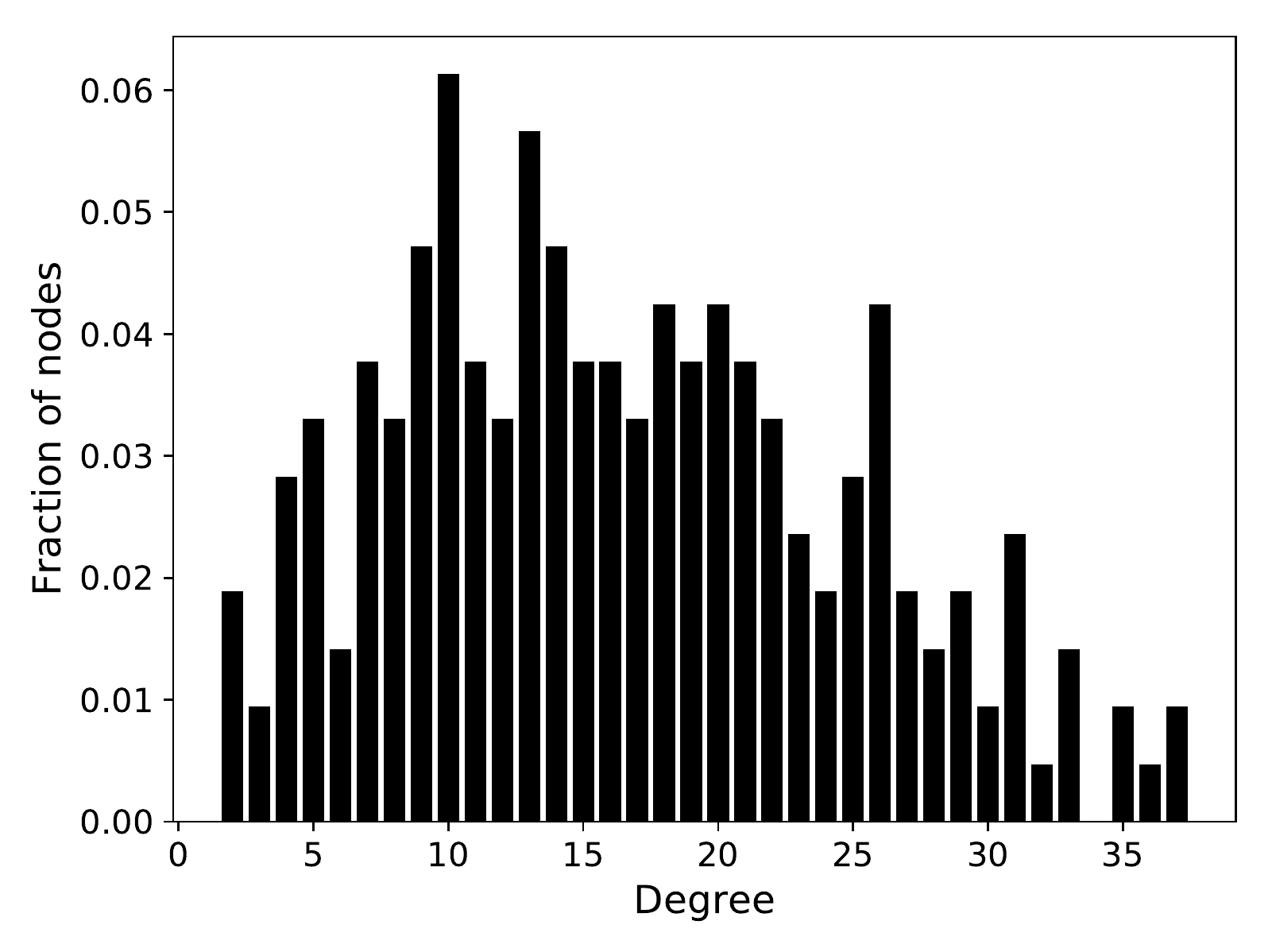}
			\includegraphics[width=0.4\textwidth]{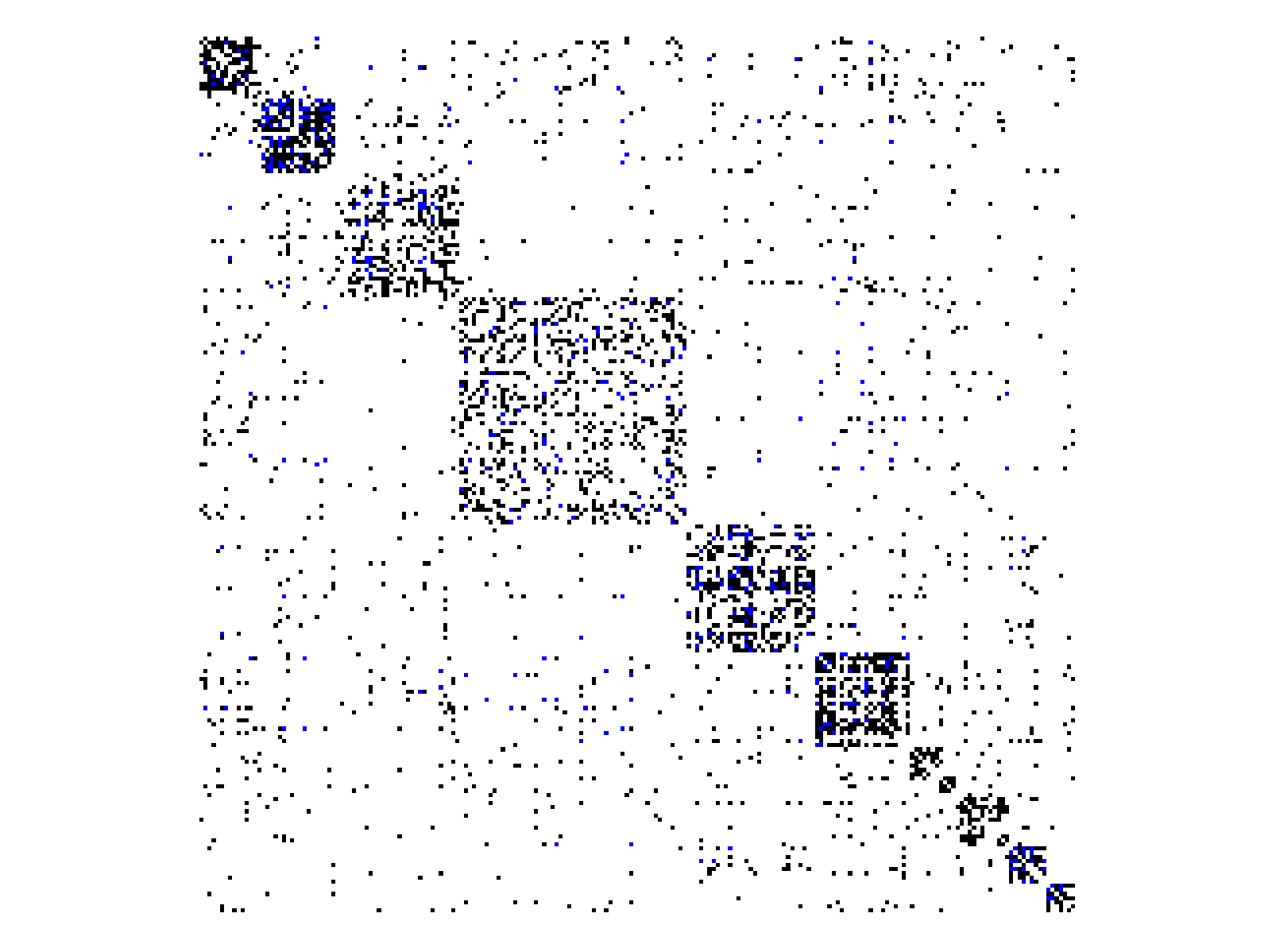}
			\subcaption{Workplace}
		\end{subfigure}\\
		\begin{subfigure}{\linewidth}
			\includegraphics[width=0.4\textwidth]{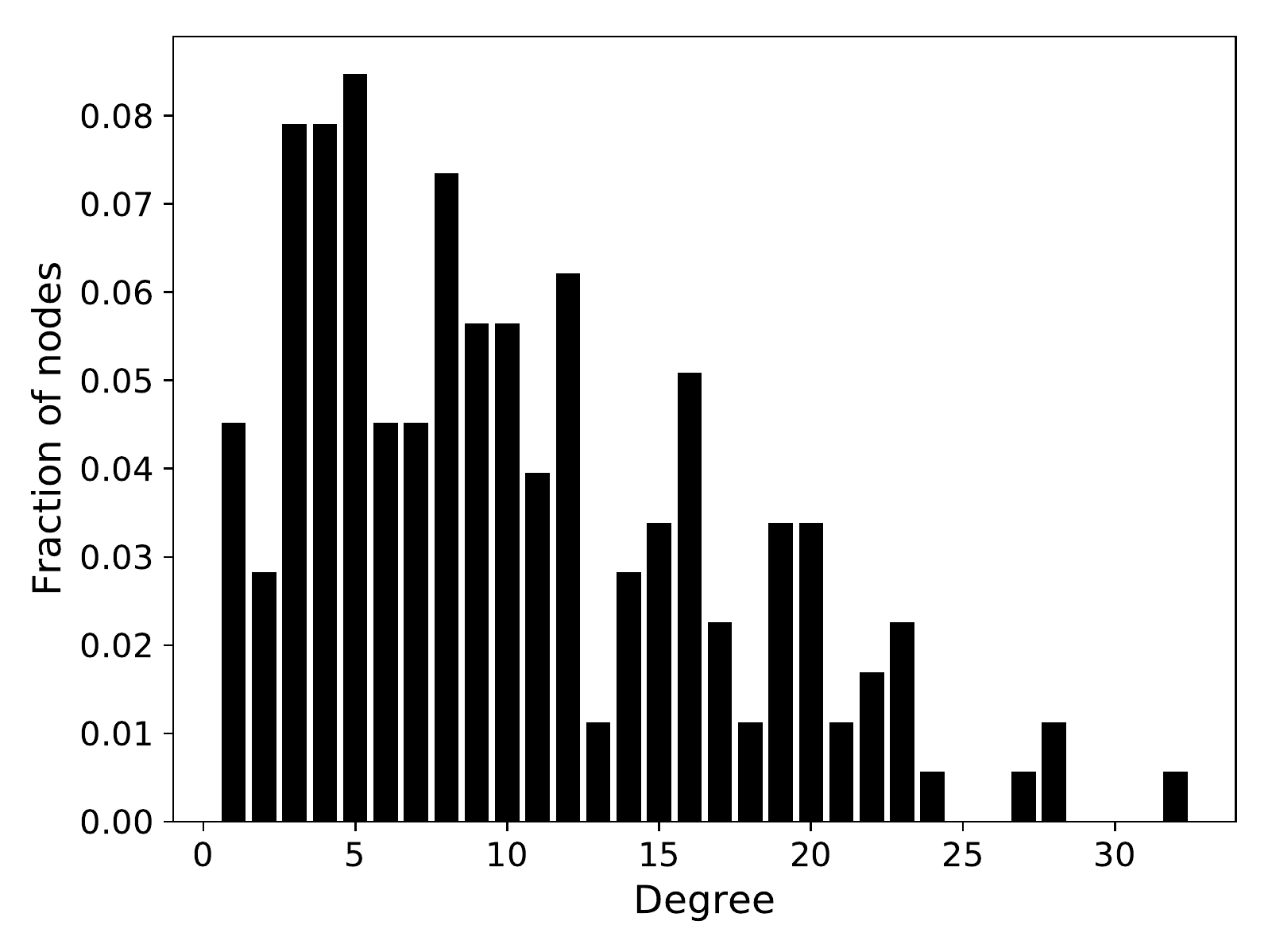}
			\includegraphics[width=0.4\textwidth]{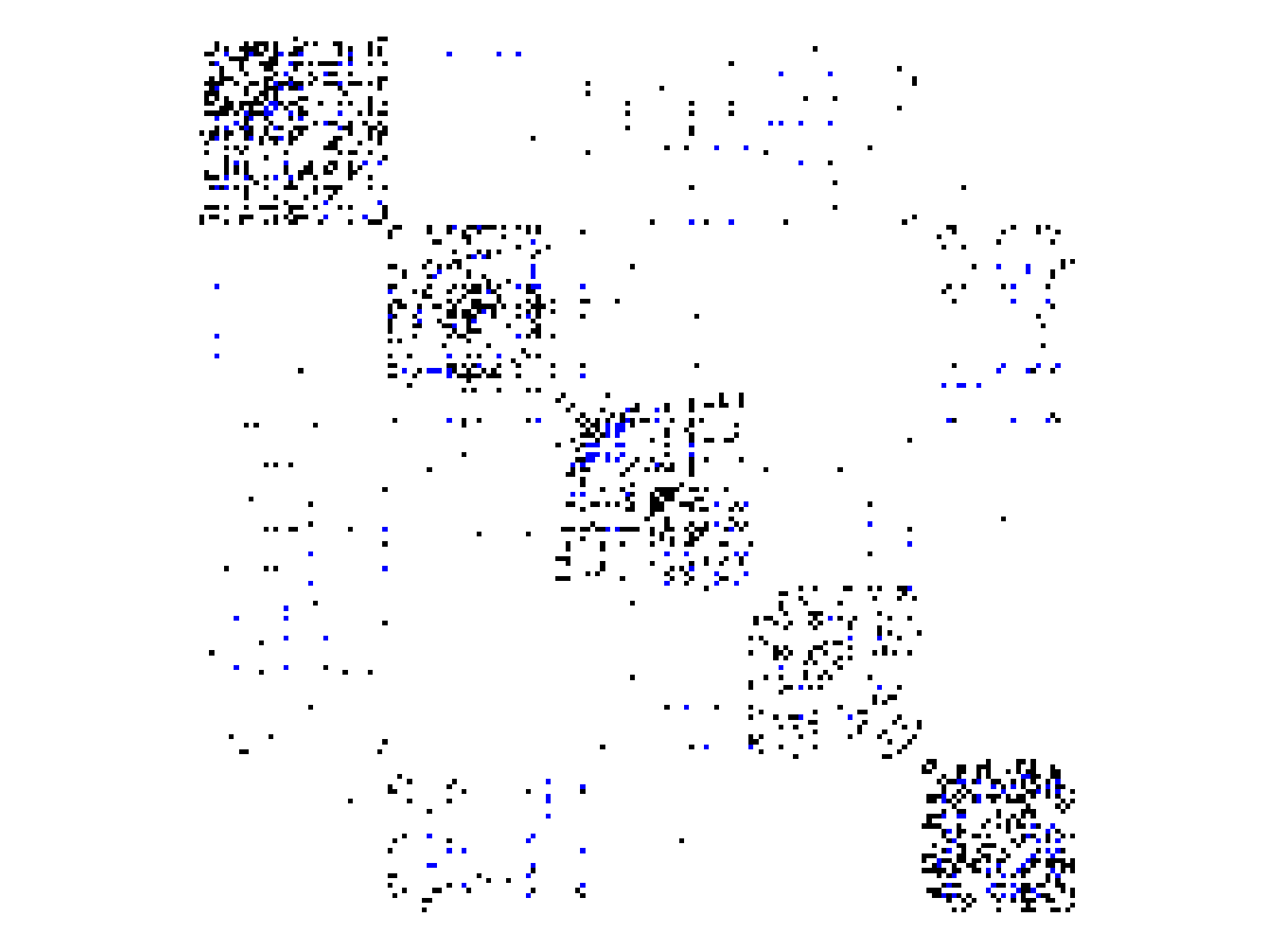}
			\subcaption{High School}
		\end{subfigure}
		\caption{Degree distribution and adjacency matrices of the hypergraphs. In the adjacency matrices, $A_{ij} = 0, 1, >1$ is coloured as white, black and blue respectively.}
		\label{fig:sociodata}
	\end{figure}

	We study the dynamics of MCM on empirical hypergraphs of social interaction.
	To create hypergraphs describing real-world interactions, we use the publicly available SocioPatterns dataset \cite{sociopatterns}. To represent a diverse variety of social situations, we use the following datasets:
	\begin{itemize}
		\item Primary Schools dataset \cite{primary_school_1}\cite{primary_school_2} - schoolchildren and teachers at a primary school in France
		\item Conference dataset \cite{sfhh_work} - participants at the 2009 SFHH conference in Nice, France
		\item Workplace dataset \cite{sfhh_work} - the staff at an office building in France (2015)
		\item High School dataset \cite{highschool} - students at a high school in Marseilles, France (2012)
	\end{itemize}
	The SocioPatterns datasets record face-to-face interactions with a temporal resolution of 20 seconds. This allows us to check whether individuals are truly interacting as a group \cite{sekara_fundamental_2016}. For every 20 second window, we create a network of interactions and catalogue all the maximal cliques. If a clique interacts often, we assume that it represents a social group where the interactions are multi-body, and are distinguishable from the pairwise interactions of its members. In practice, we set an arbitrary threshold and include a maximal clique as a hyperedge if it is observed $\geq 5$ times.
	Specifics about the structure of the hypergraphs are provided in Table \ref{tab:socio}. Fig. \ref{fig:sociodata} shows their degree distributions and weighted adjacency matrices. Here, degree refers to the number of hyperedges containing a node and $A_{ij}$ is given by the number of hyperedges containing both $i$ and $j$.
	
	\begin{table}
		\centering
		{\renewcommand{\arraystretch}{1.5}
			\begin{tabular}{|c|c|c|cccccccc|}
				\hline
				\textbf{Hypergraph}& \textbf{$N$} & \textbf{$M$} & $m_2$ & $m_3$ & $m_4$ & $m_5$ & $m_6$ & $m_7$ &  $m_8$ & $m_9$ \\
				\hline
				Primary School & 242 & 3463 & 3118 & 338 & 7 & 0 & 0 &0 &0 &0 \\
				Conference & 392 & 1710 & 1441 & 224 & 29 & 8 & 3 & 2 & 2 & 1\\
				Workplace & 212 & 1703 & 1606 & 97 & 0 & 0 & 0 & 0 & 0 & 0 \\
				High School & 177 & 866 & 795 & 69 & 2 & 0 & 0 & 0 & 0 & 0\\
				\hline
		\end{tabular}}
		\caption{Details of hypergraphs created from real-world data - number of nodes $N$, number of hyperedges $M$, number of $k$-edges $m_k$}
		\label{tab:socio}
	\end{table}
	
	The resulting hypergraphs show very different types of structures, including different degree distributions. This diversity is an asset in order to  test the qualitative features of the dynamics of MCM.
	
	\subsection{Role of involvement}
\begin{figure}
		\centering
		\begin{subfigure}{\linewidth}
			\includegraphics[width=0.3\textwidth]{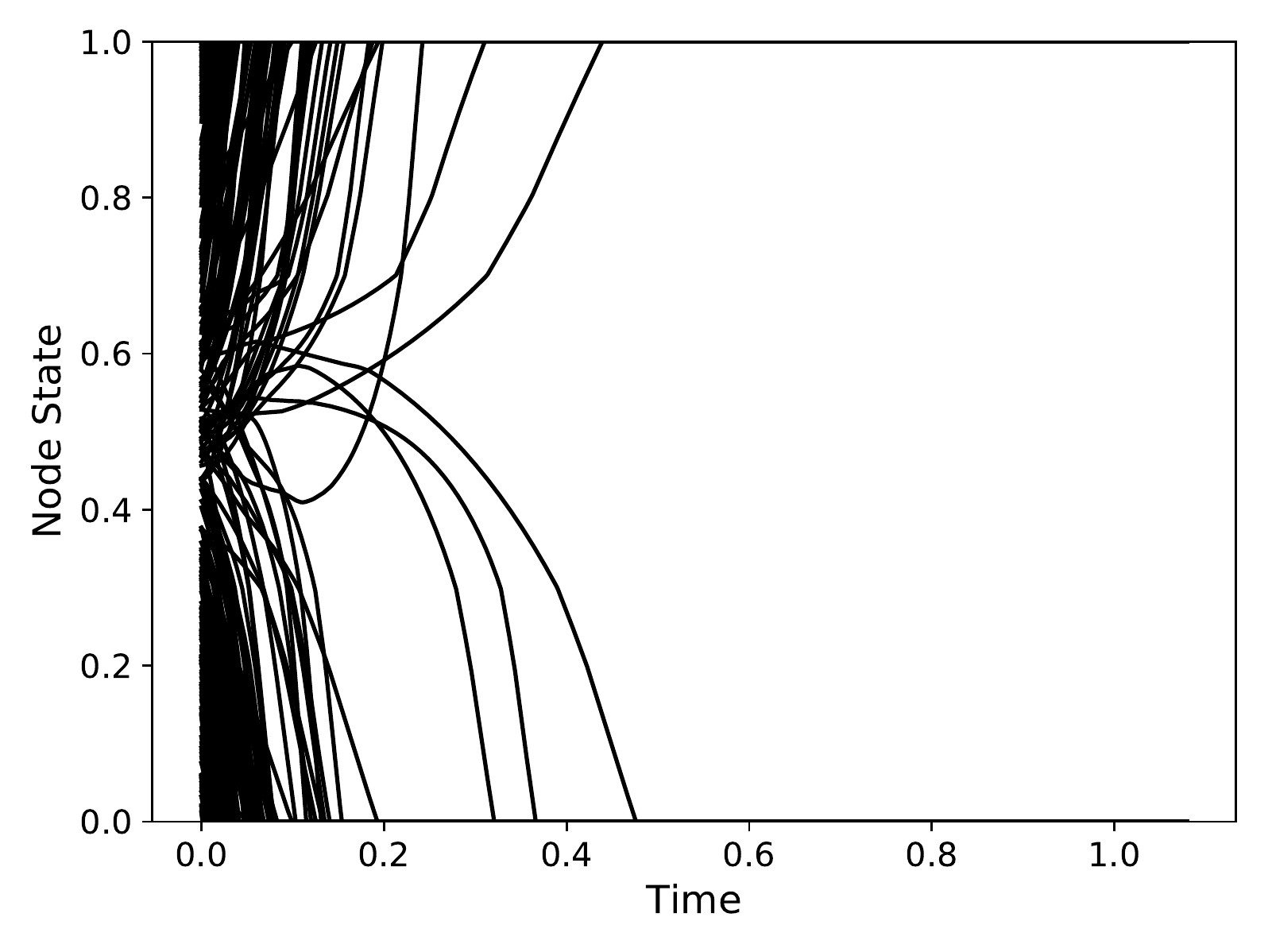}    
			\includegraphics[width=0.3\textwidth]{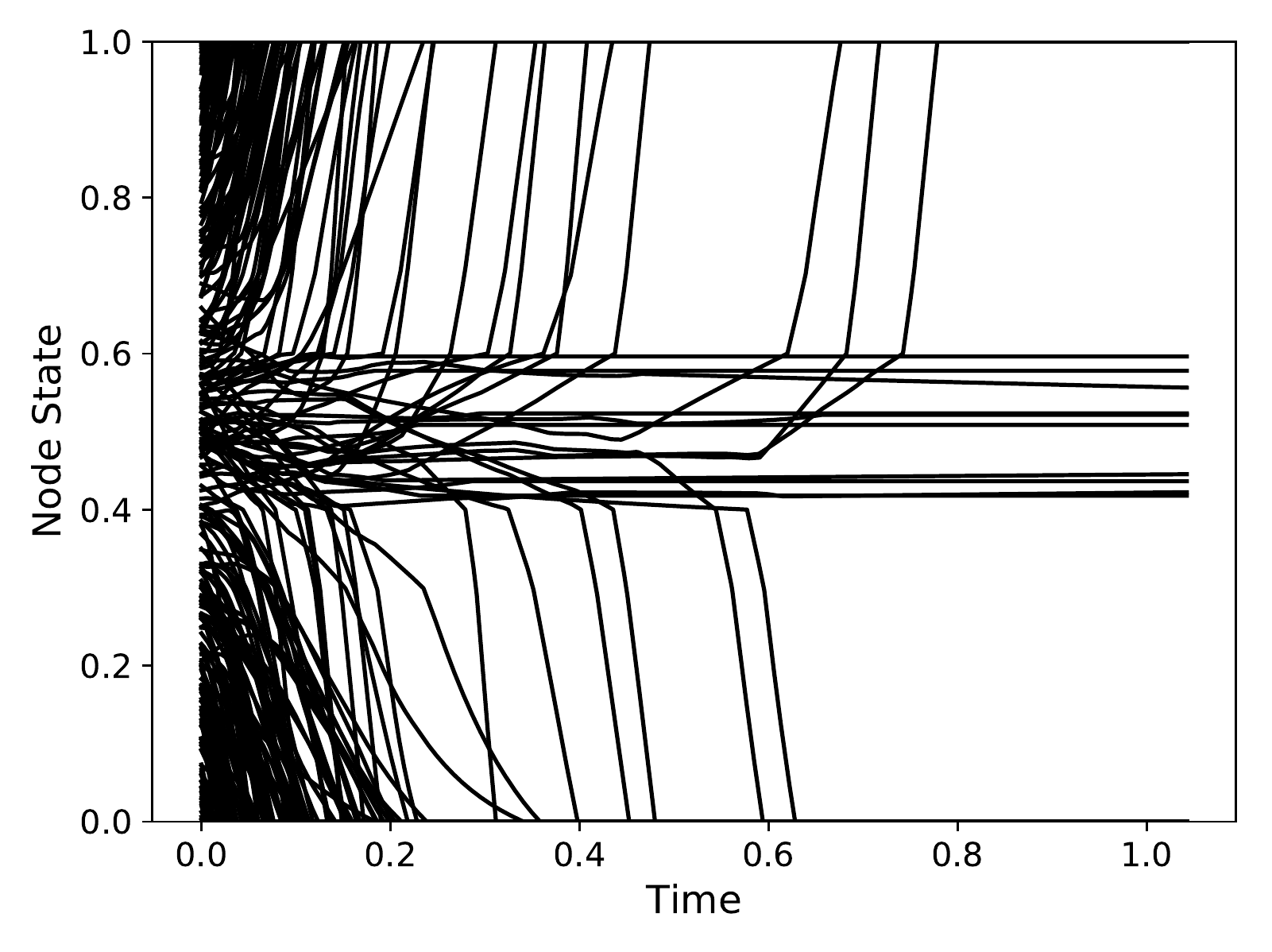}
			\includegraphics[width=0.3\textwidth]{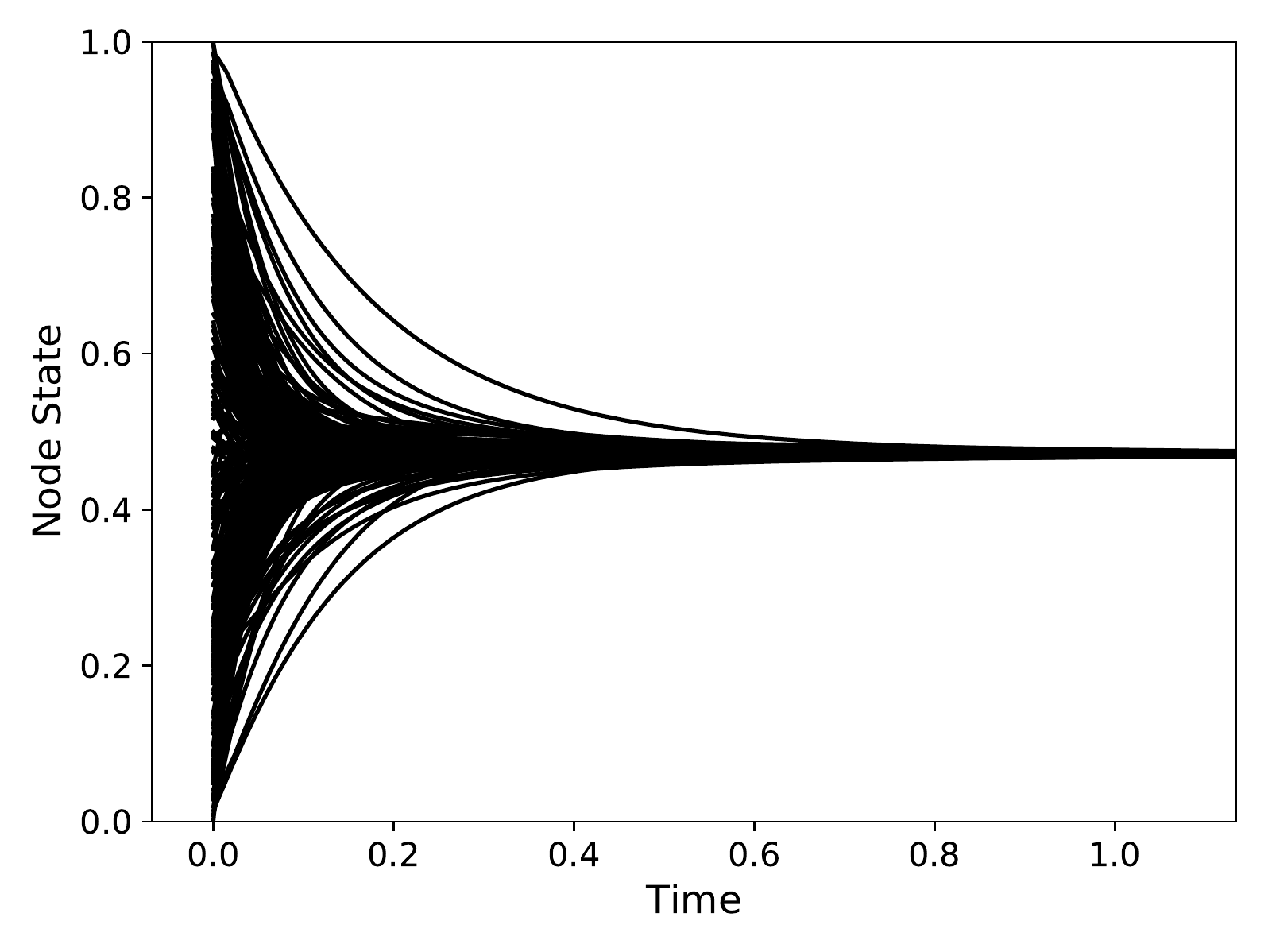}
			\subcaption{Primary School}		
		\end{subfigure}	\\
		\begin{subfigure}{\linewidth}
			\includegraphics[width=0.3\textwidth]{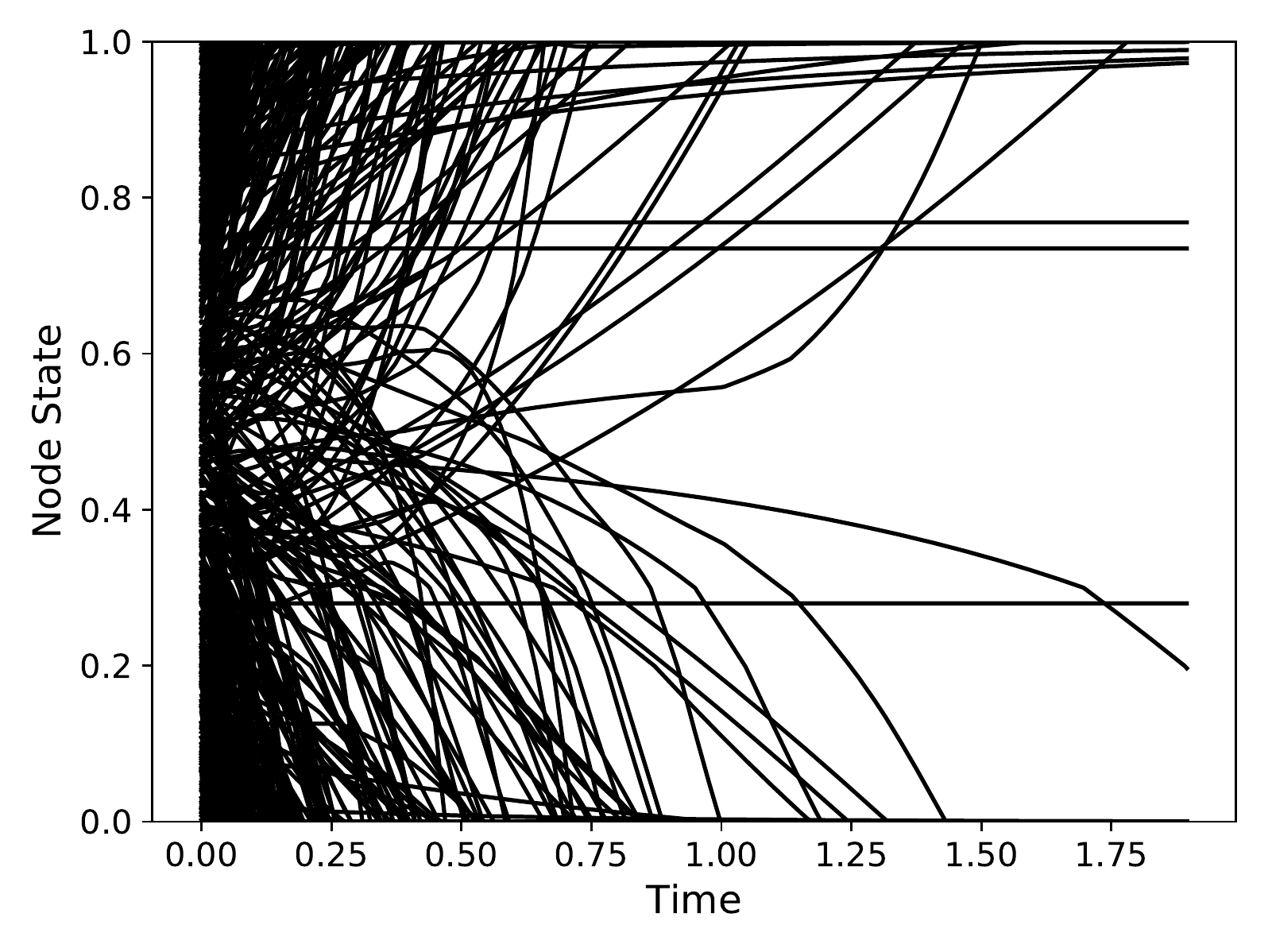}    
			\includegraphics[width=0.3\textwidth]{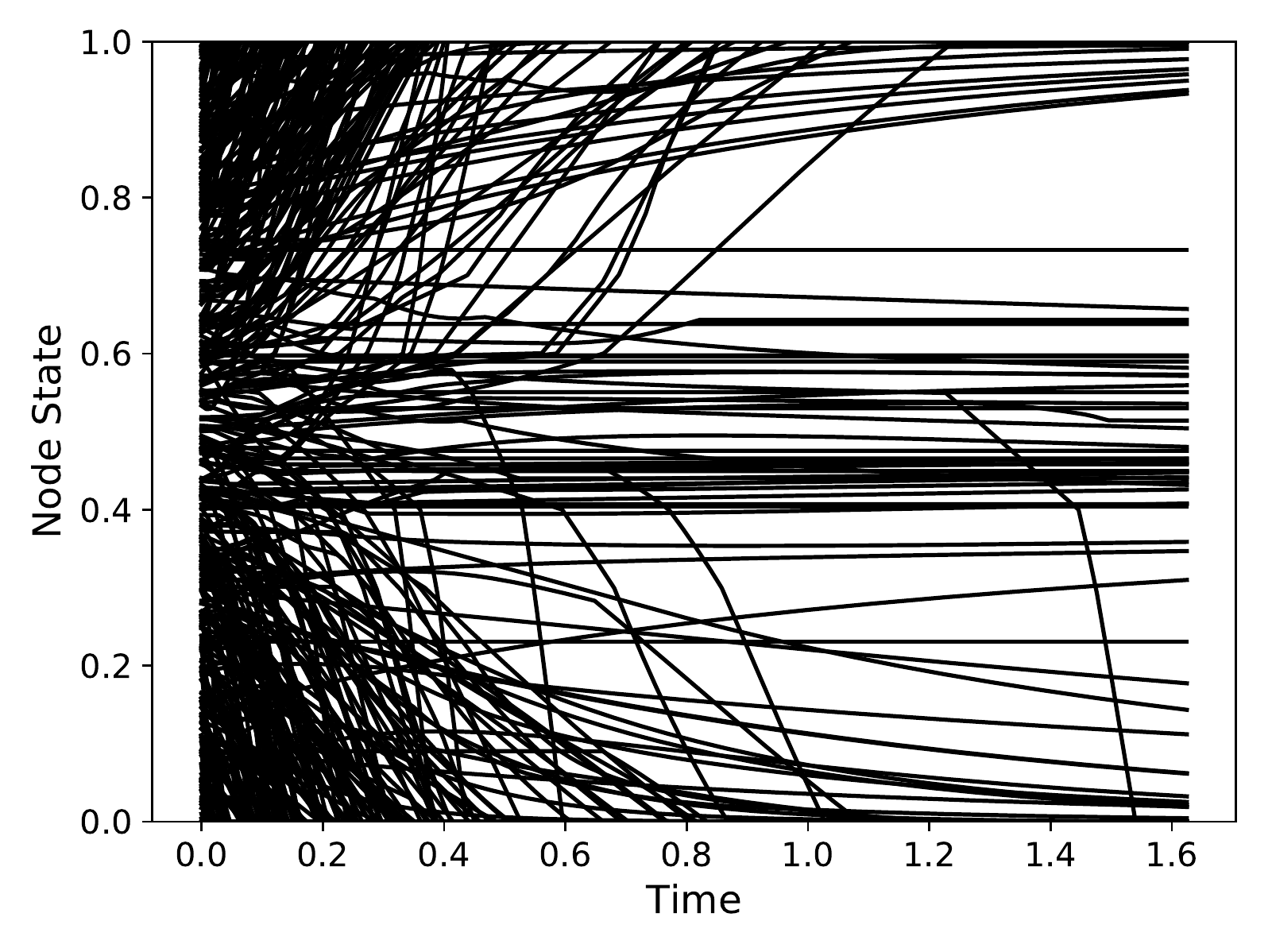}
			\includegraphics[width=0.3\textwidth]{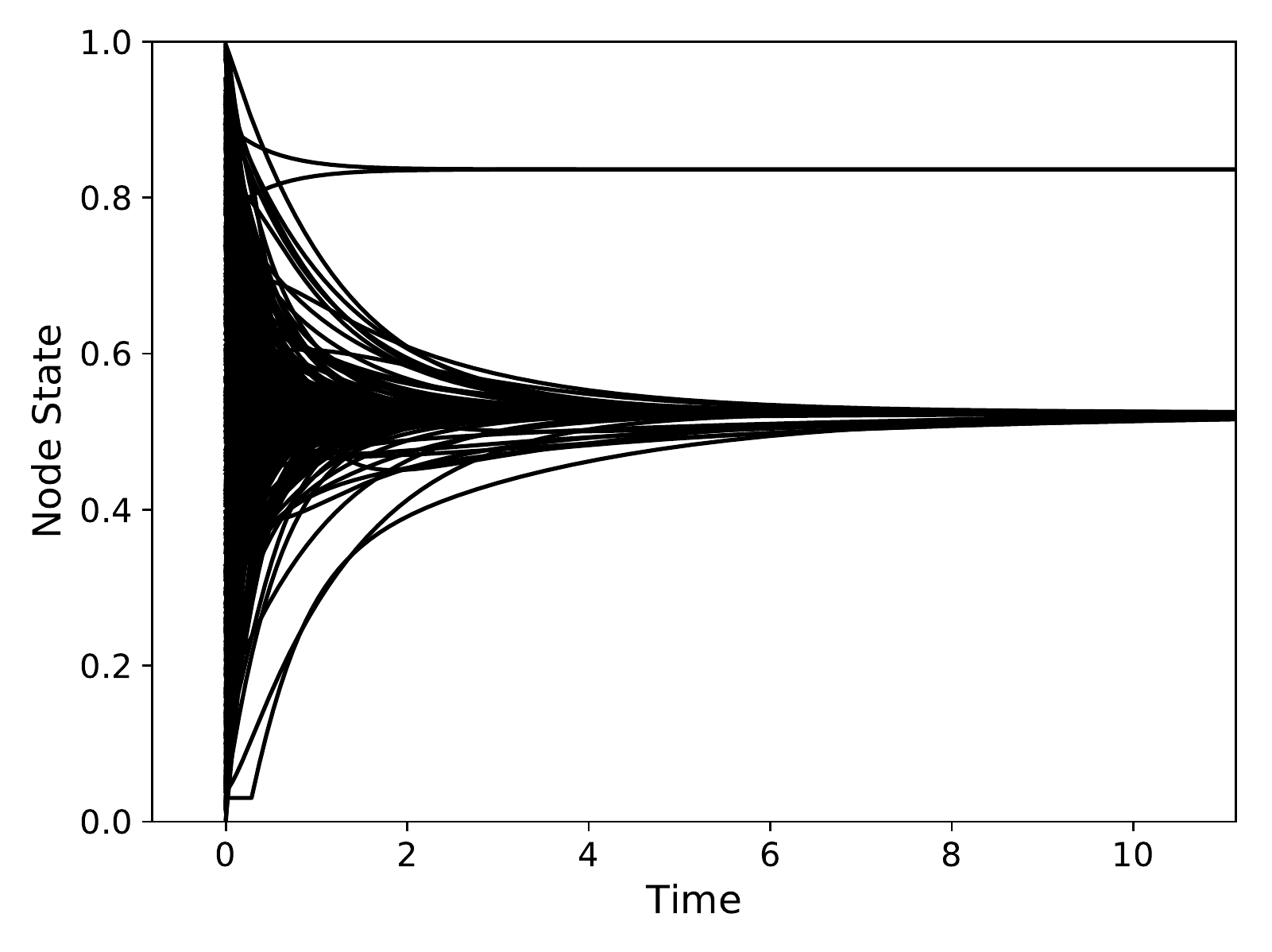}
			\subcaption{Conference}		
		\end{subfigure}	\\   
		\begin{subfigure}{\linewidth}
			\includegraphics[width=0.3\textwidth]{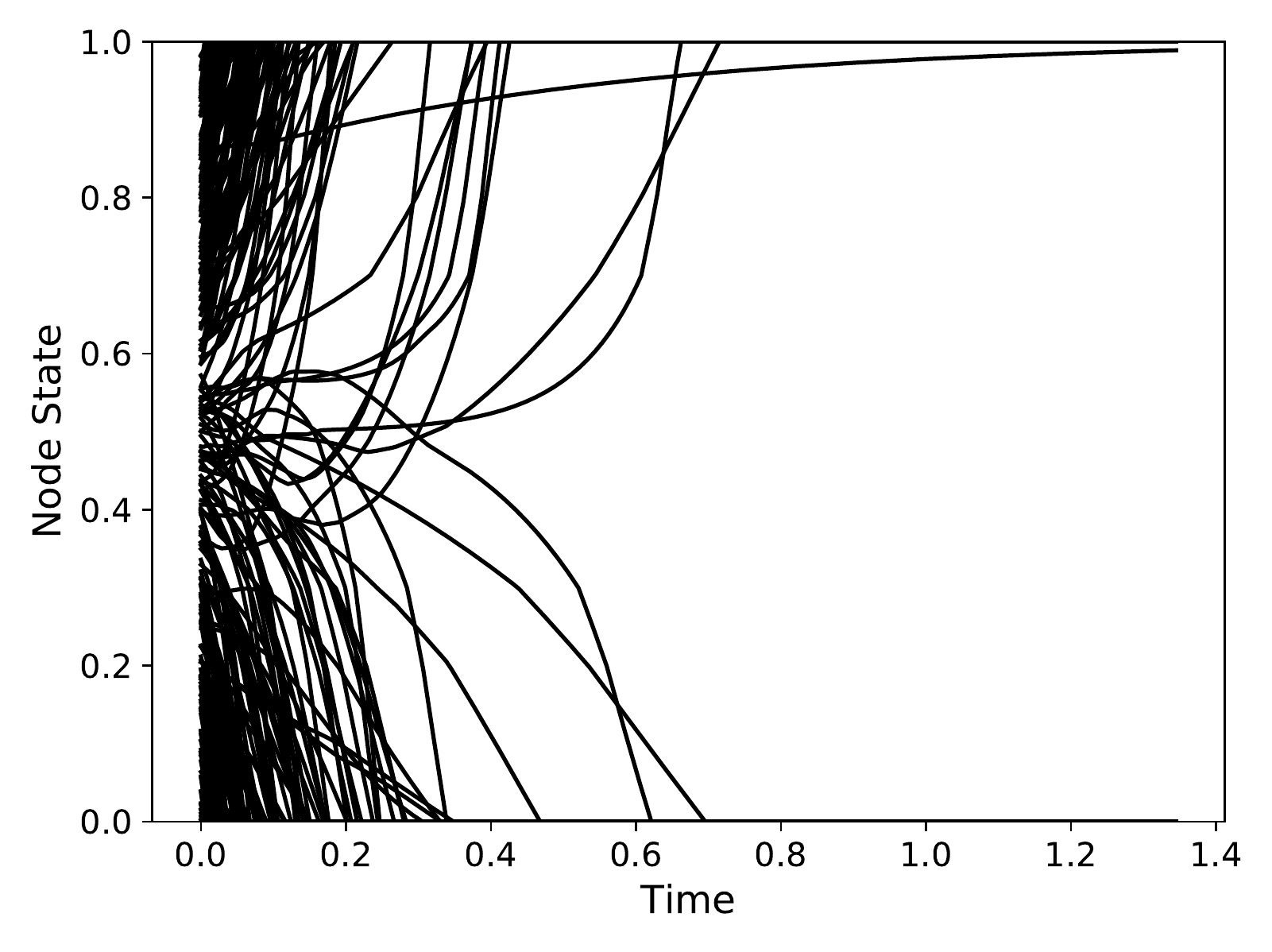}    
			\includegraphics[width=0.3\textwidth]{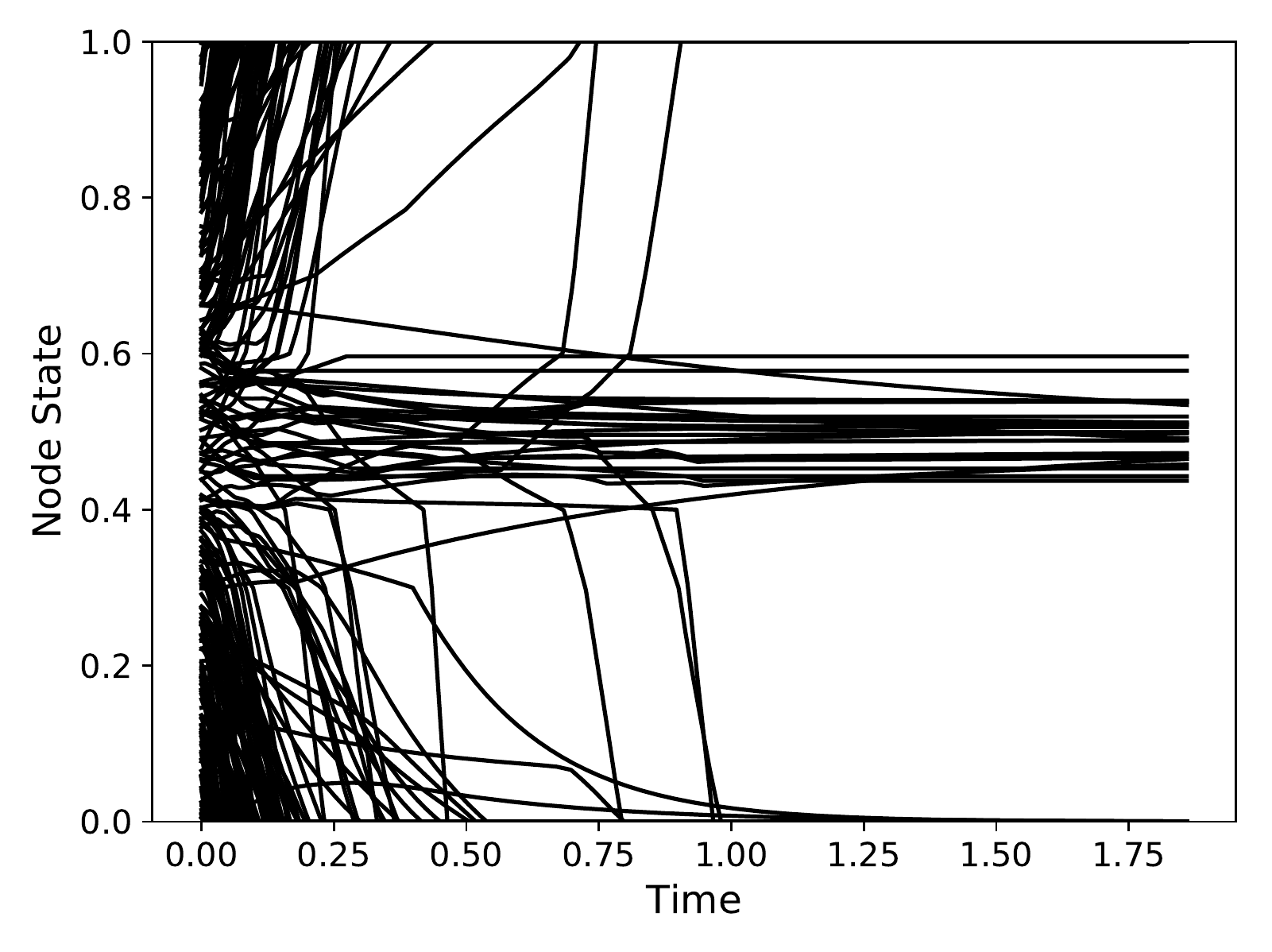}
			\includegraphics[width=0.3\textwidth]{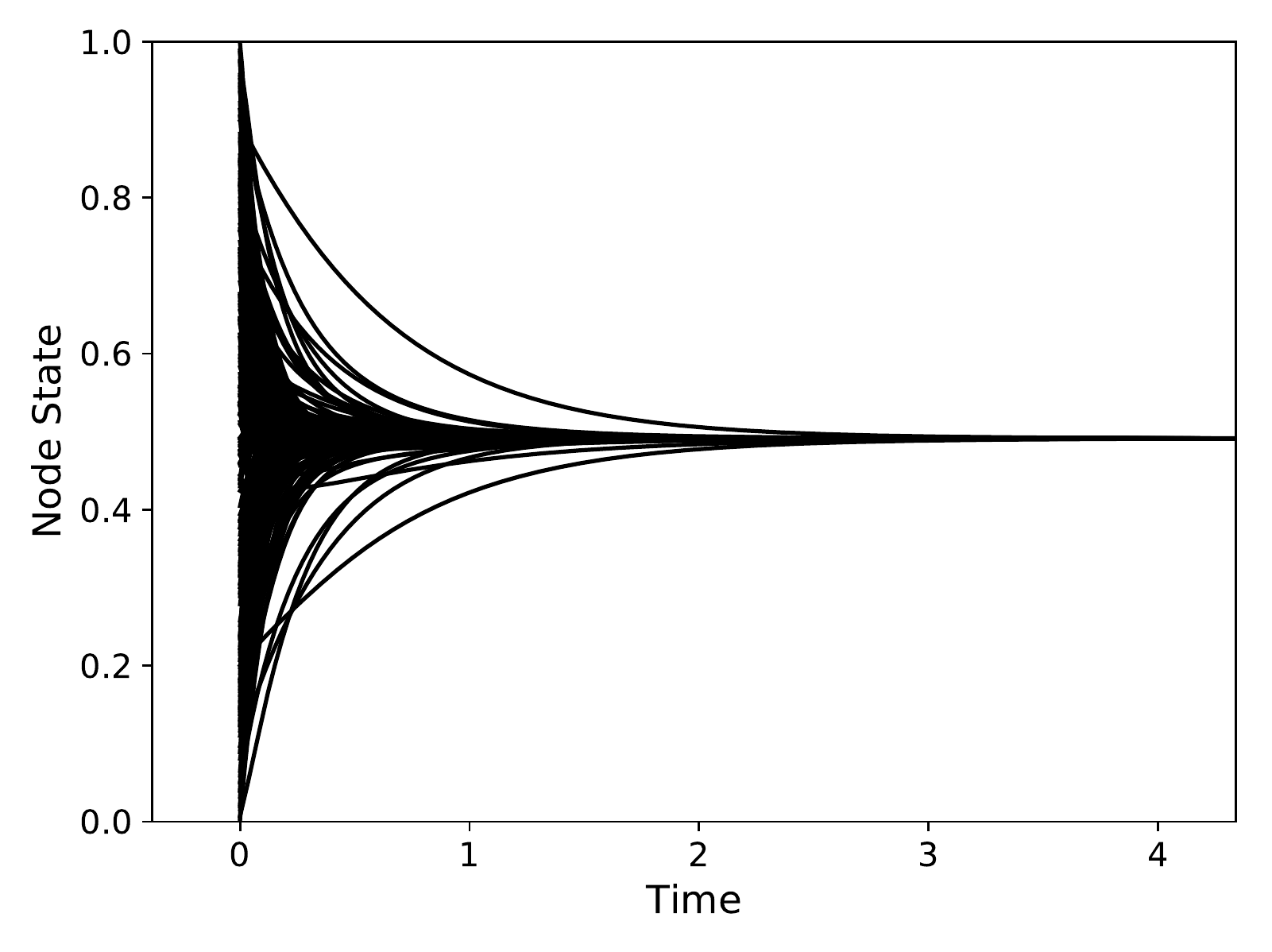}
			\subcaption{Workplace}		
		\end{subfigure}\\
		\begin{subfigure}{\linewidth}
			\includegraphics[width=0.3\textwidth]{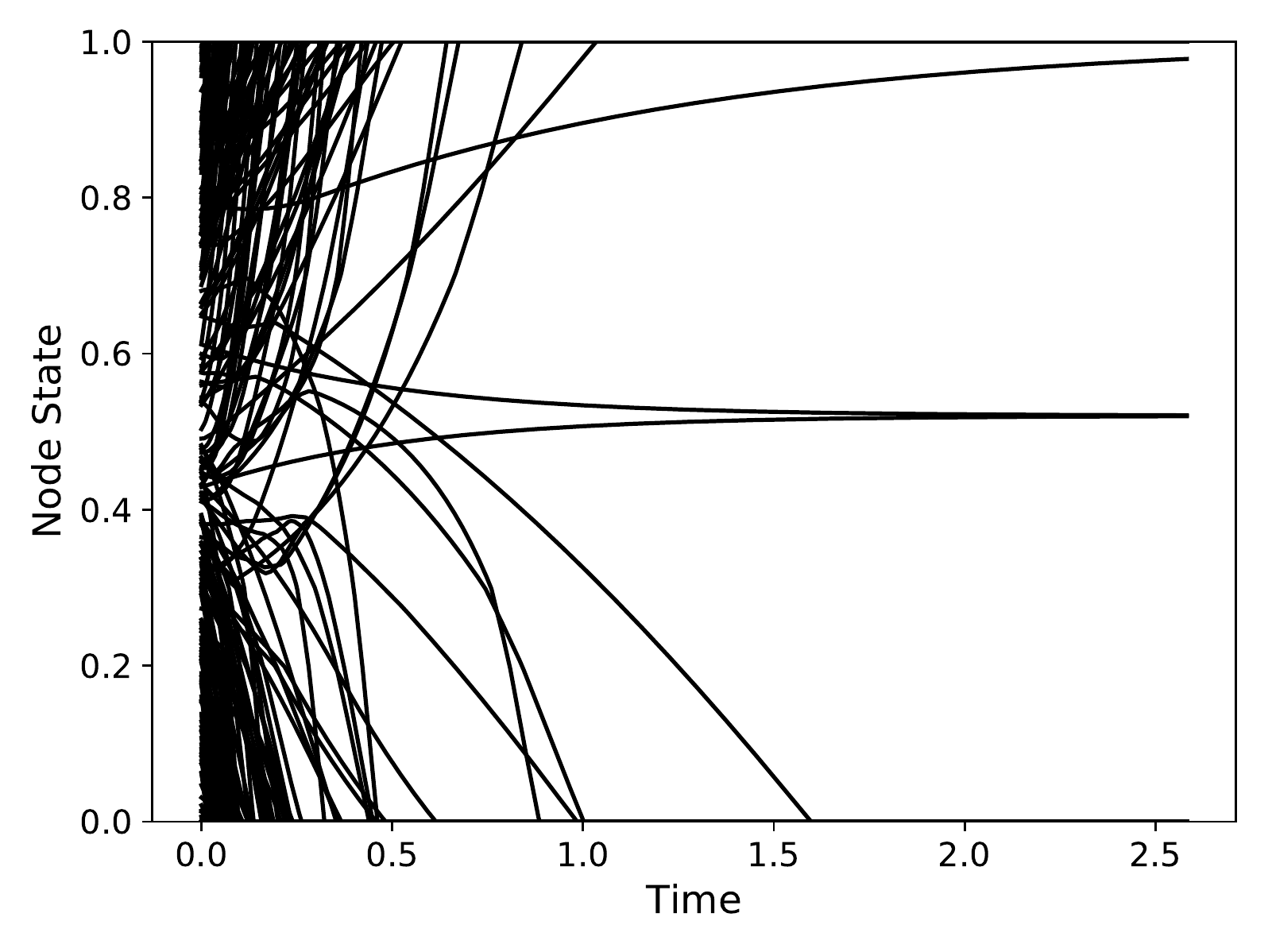}    
			\includegraphics[width=0.3\textwidth]{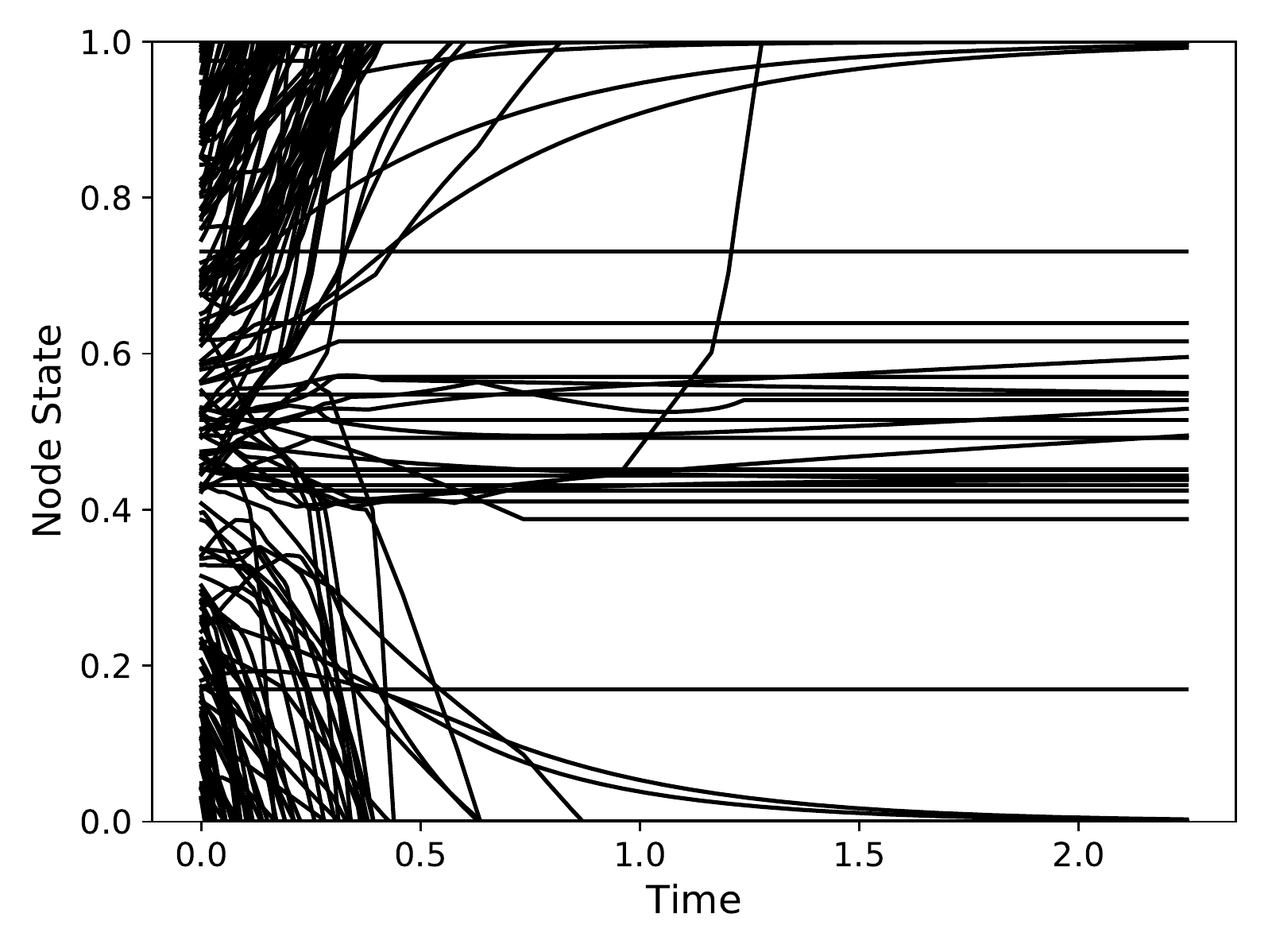}
			\includegraphics[width=0.3\textwidth]{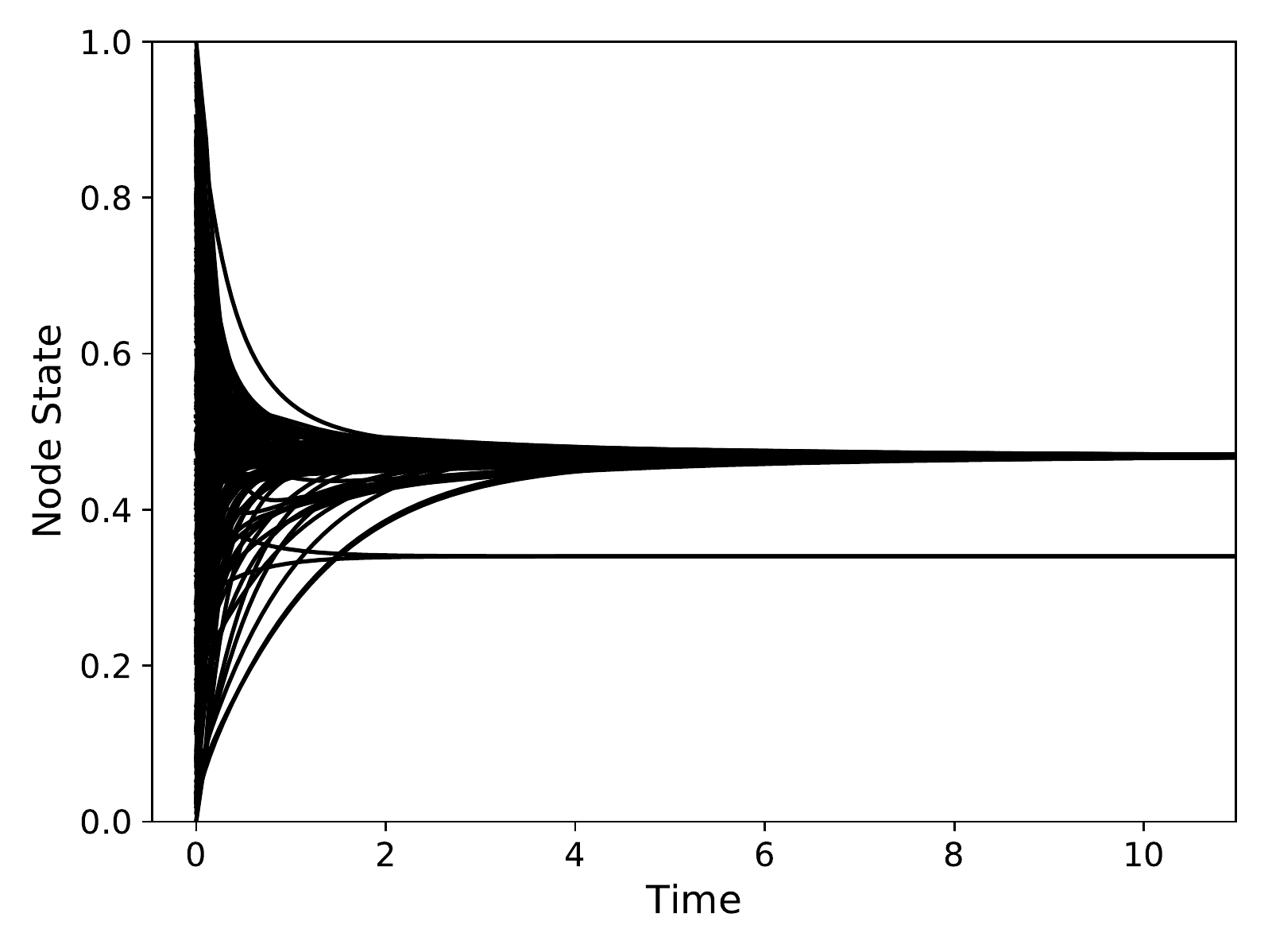}
			\subcaption{High School}		
		\end{subfigure}	\\	 
		\begin{subfigure}{\linewidth}
			\includegraphics[width=0.3\textwidth]{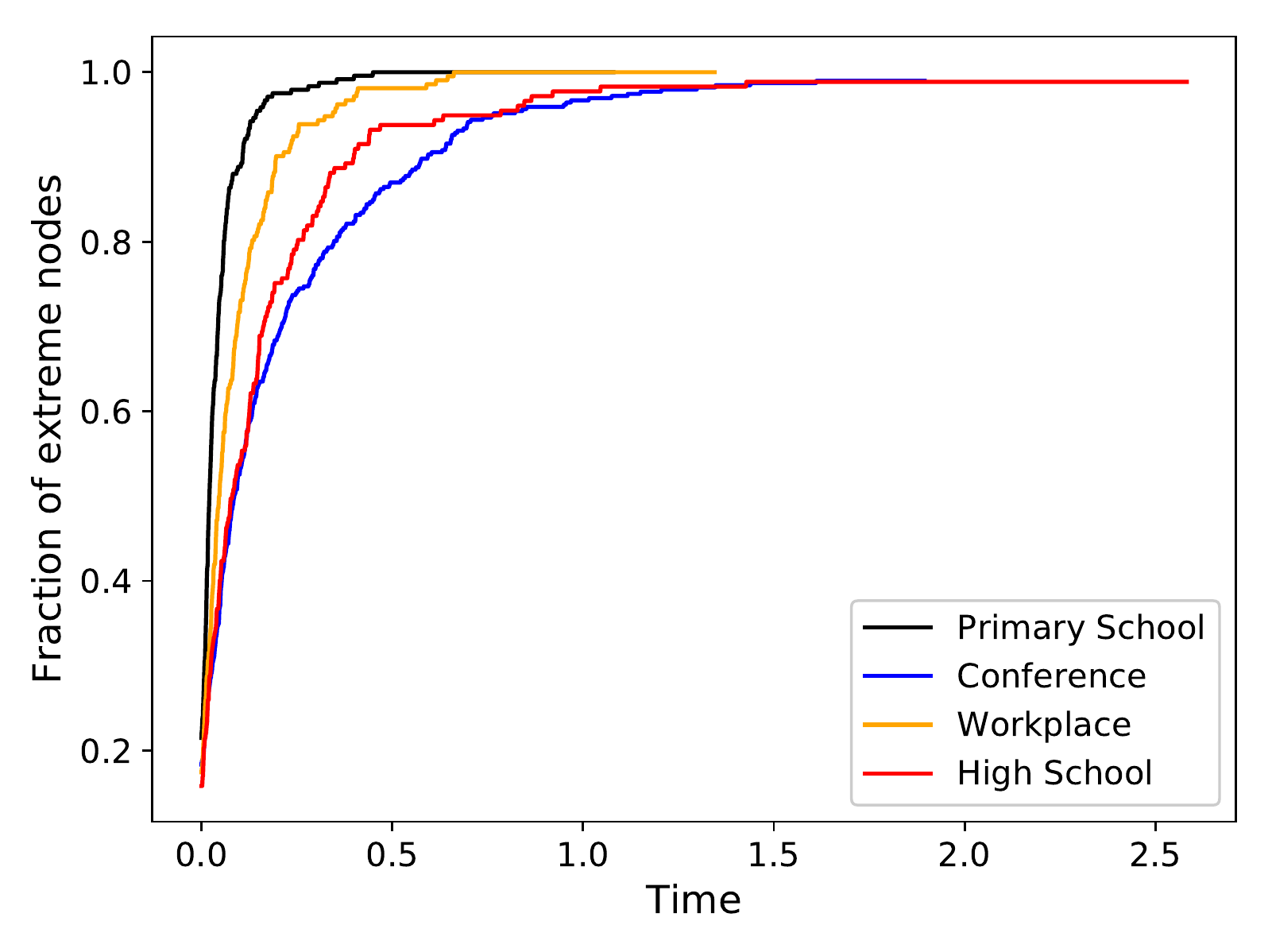}    
			\includegraphics[width=0.3\textwidth]{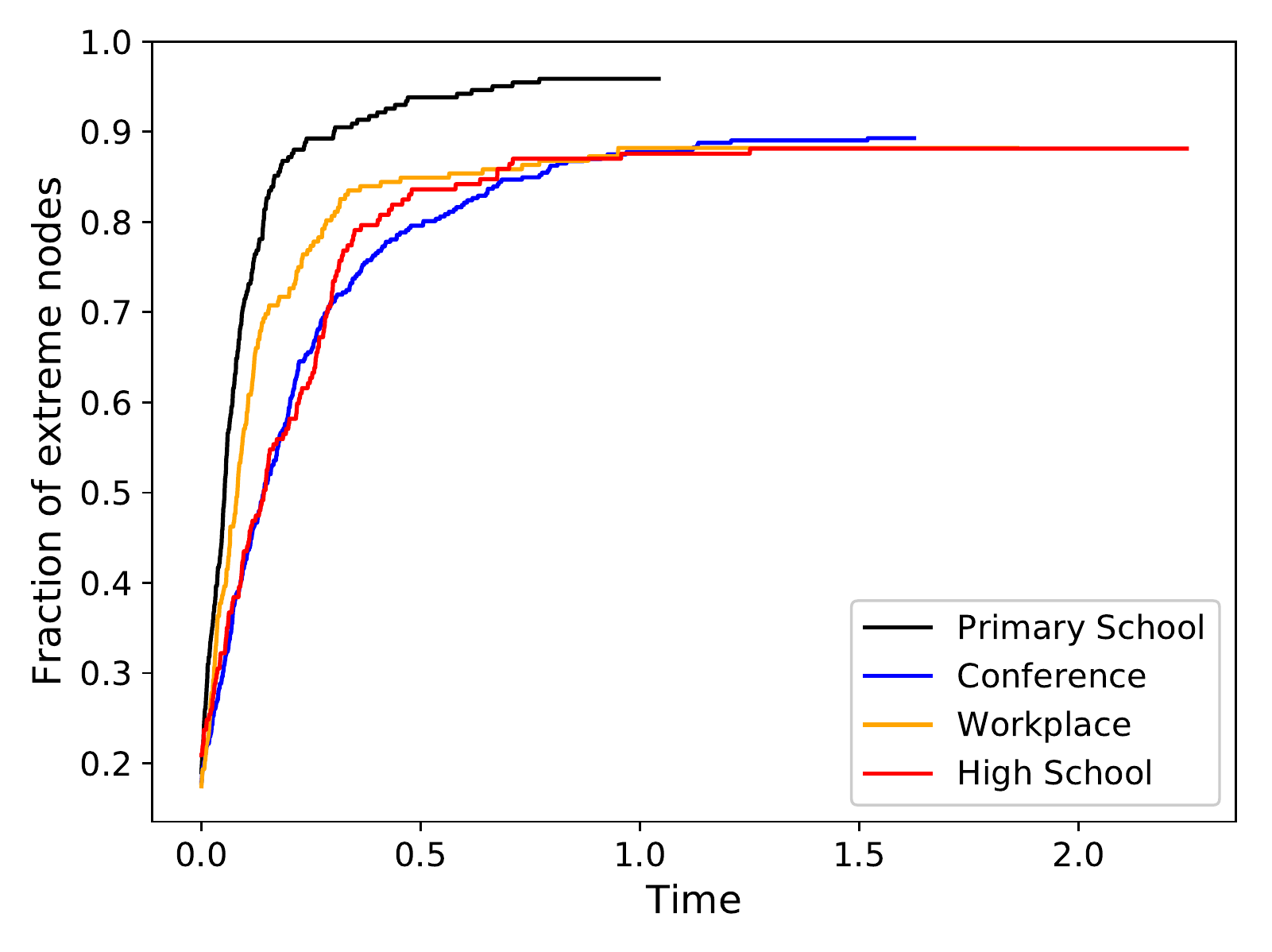}
			\includegraphics[width=0.3\textwidth]{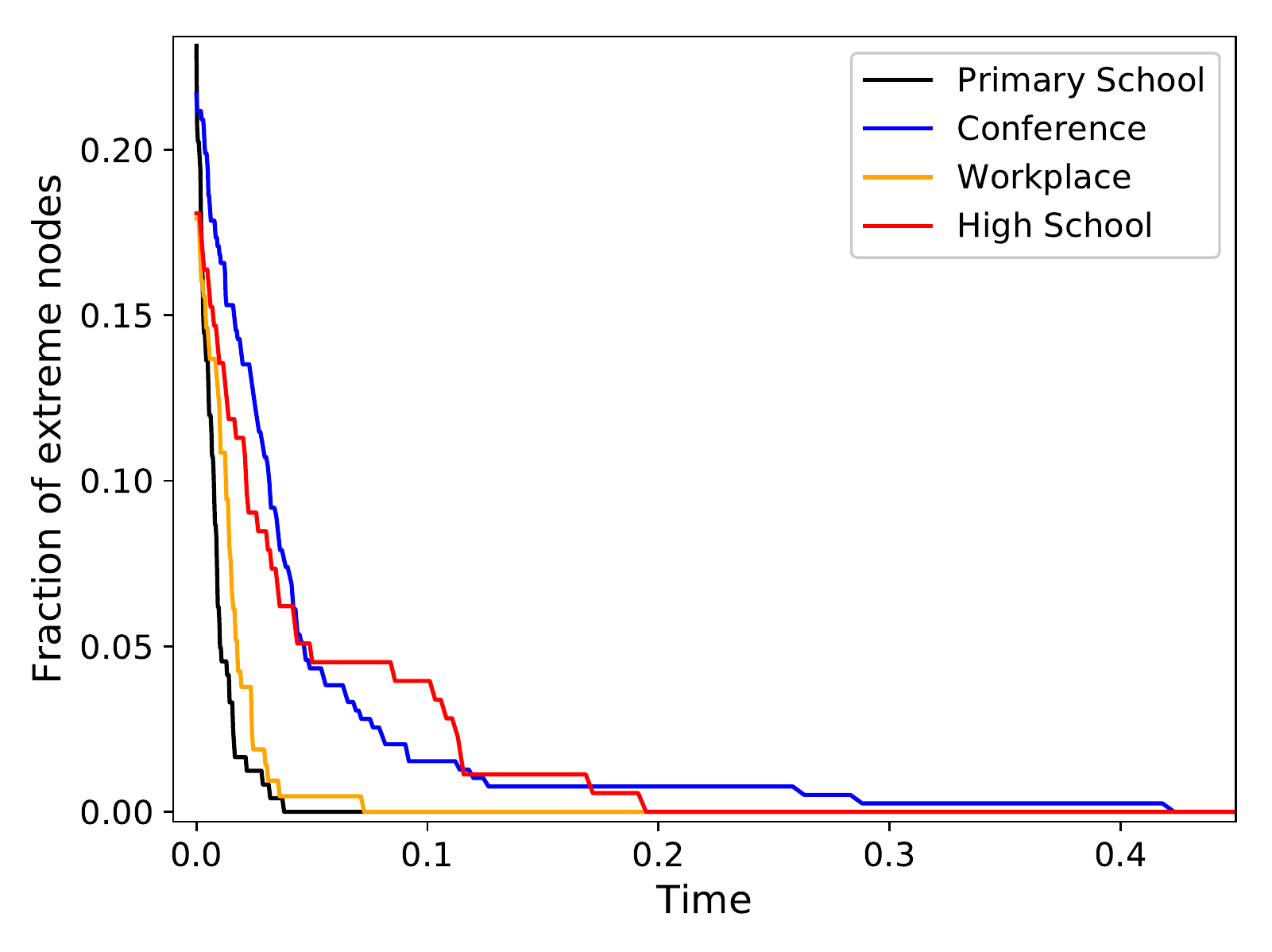}
			\subcaption{Fraction of extreme nodes}		
		\end{subfigure}	
		\caption{Polarisation in High (left), medium (center) and low (right) involvement scenarios.}
		\label{fig:polarisation}
	\end{figure}

	We initialise the nodes with states drawn from a uniform distribution in $[0,1]$. Fixing $\lambda_i = -1$, $\delta_i = -5$, we consider three scenarios for (\ref{oppo}):
	\begin{itemize}
		\item High involvement - $\phi^A_i=0.10$, $\phi^R_i=0.15$
		\item Medium involvement - $\phi^A_i=0.15$, $\phi^R_i=0.30$
		\item Low involvement - $\phi^A_i=0.40$, $\phi^R_i=0.80$
	\end{itemize}
	in order to investigate the role of involvement on the dynamics. Note that the goal of this investigation is to show that there are values of the parameters for which the model evolves in similar ways for all the empirical hypergraphs. A complete exploration of the parameter space is beyond the intent and the scope of this work.
		Figs. \ref{fig:polarisation} (a) - (d) show the typical evolution of one realisation of MCM on the empirical hypergraphs for each scenario. 
	To quantify the extent of polarisation, we plot the fraction of extreme nodes in the course of time (Fig. \ref{fig:polarisation} (e)).
	Fig. \ref{fig:polarisation} shows that, at least for our choice of parameters, the evolution of MCM is similar in the different topologies, indicating the robustness of temporal profiles on the details of the underlying hypergraph. We observe that a group of individuals who are all highly invested in a discussion tends to split into two extreme factions with opposing views. Groups with less invested individuals tend to settle on a common consensus, while groups of moderately invested people evolve to an intermediate situation, with a partial consensus and the emergence of a small group of extremists. 
	
	\subsection{Role of stubbornness}
\begin{figure}
		\centering
		\begin{subfigure}{0.45\linewidth}
			\includegraphics[width=\textwidth]{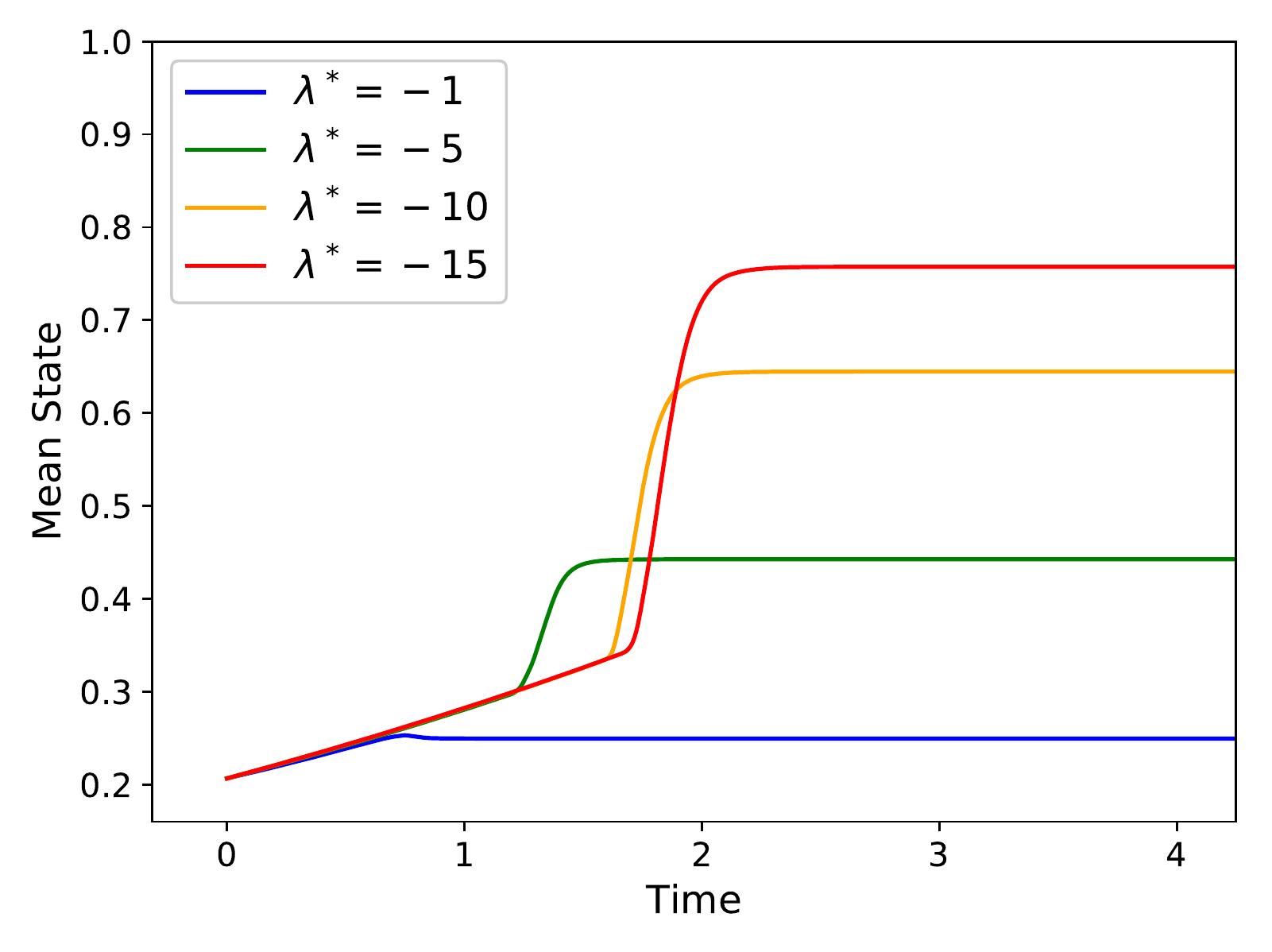}
		\end{subfigure}
		\begin{subfigure}{0.45\linewidth}
			\includegraphics[width=\textwidth]{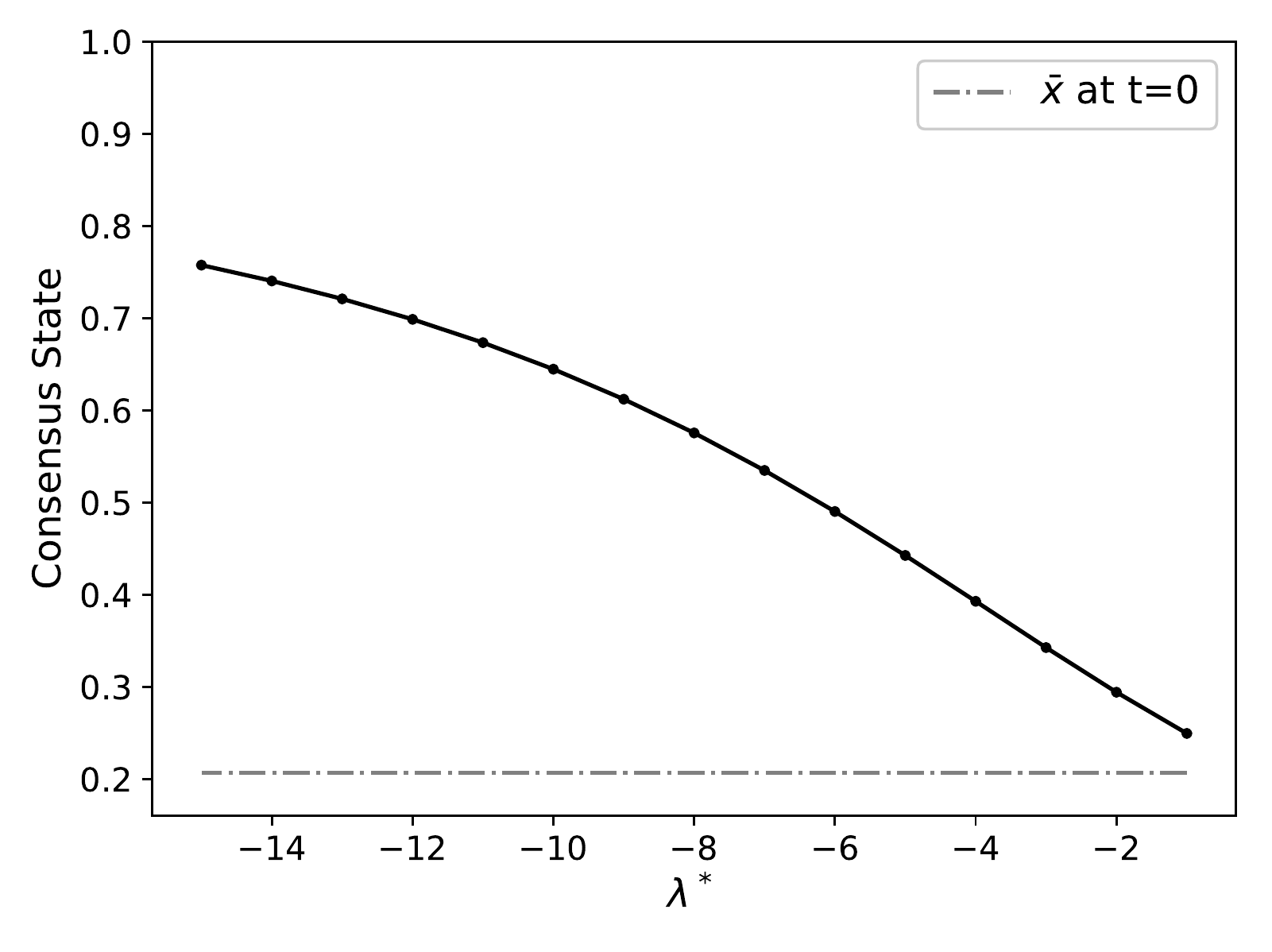}
		\end{subfigure}
		\caption{Stubbornness in the Primary Schools hypergraph (a) evolution of the mean state with time and (b) consensus state vs $\lambda^*$}	
		\label{fig:stub_ps}
	\end{figure}

	As a second step, we investigate the impact of stubbornness. To facilitate the understanding of the results, we restrict the scope to situations where all nodes are subject solely to acceptance, and not rejection, by imposing that $\phi_A , \phi_R > 1$. We also fix the same value of $\delta_i $ for all the nodes, with $\delta_i = -5$ in the following simulations, but qualitatively similar results can be found for other values. The value of $\lambda_i$ defines the stubbornness for each node $i$, such that $\lambda_i < 0$ characterises a stubborn node and stubbornness increases with the magnitude of $\lambda_i$. Numerical simulations reveal that the effect of changing $\lambda_i$ is sensitive to the topology, initialisation as well as the values of the other parameters. To fix the topology, we consider the Primary School hypergraph.
	Fig. \ref{fig:sociodata} shows that the Primary School hypergraph is reminiscent of a block model. Students interact much more with their classmates than with others.  In order to ensure that the effect of stubborness is present in  every block, we fix the stubborn nodes - choosing all $10$ teachers and picking $3$ students randomly from each class ($50$ in total).	
	We initialise all the nodes with binary states - $1$ for the stubborn nodes ($\lambda_i= \lambda^*$), and $0$ for the others ($\lambda_i = -1$). This gives us $\bar{x}(t=0) = 0.2066$. The results in Fig. \ref{fig:stub_ps} show that the mean opinion of the group shifts towards $1$, which indicates that the presence of a few stubborn individuals has a significant effect on the consensus of the group. Further, the dependence of the consensus state on the value of $\lambda^*$ shows that the effect of stubborn individuals increases with their stubbornness.

	\section{Discussion}\label{sec:discussion}
	Networks are a powerful language to model interacting systems. Yet one of its core assumptions, that connectivity emerges from the combination of pairwise interactions, is not necessarily verified in empirical systems. In this paper, we have explored the problem of consensus in  situations such that basic interaction units are composed of more than two nodes. Representing the underlying structure as a hypergraph, we propose a general non-linear model for consensus dynamics, called MCM, where different model ingredients are associated with different sociological mechanisms.  We have studied certain aspects of the models mathematically and have explored its behaviour on artificial as well as on real-life hypergraphs, revealing a rich phenomenology and the strong interplay between the structure and the dynamics.
	Yet, this paper only scratches the surface of this dynamical system, and many research venues remain open.  First, we have explored only a limited part of the parameter space, and additional theoretical and numerical results would be required to improve our understanding of the different regimes supported by the model and the nature of the transitions between them.
	Second, though we define multi-body interactions in a general way, we analyse only a special class of them. Finding different functional forms of group interactions relevant to other complex processes is a possible direction of research.
	Another interesting line of inquiry lies in extending and generalising multi-body interactions to adaptive hypergraphs. Most dynamical processes (including MCM) are sensitive to the topology, and thus making models to capture the dynamics of the underlying hypergraph promises to be an important area.
	Other possible extensions could be studying opinion dynamics on weighted or directed hypergraphs, or exploring the effects of stochasticity on the consensus.
	
	\section{Acknowledgements}
	RS is supported by the DST-Inspire scholarship.
	LN would like to thank the Hertie School for financial support.
	
	\bibliographystyle{ieeetr}
	\bibliography{references}

\begin{thebibliography}{10}

\bibitem{social_dyn_review}
C.~Castellano, S.~Fortunato, and V.~Loreto, ``Statistical physics of social
  dynamics,'' {\em Reviews of Modern Physics}, vol.~81, p.~591–646, Nov 2009.

\bibitem{DeGroot1974}
M.~H. DeGroot, ``Reaching a consensus,'' {\em Journal of the American
  Statistical Association}, vol.~69, no.~345, pp.~118--121, 1974.

\bibitem{deffuant2000mixing}
G.~Deffuant, D.~Neau, F.~Amblard, and G.~Weisbuch, ``Mixing beliefs among
  interacting agents,'' {\em Advances in Complex Systems}, vol.~3, no.~01n04,
  pp.~87--98, 2000.

\bibitem{watts2002simple}
D.~J. Watts, ``A simple model of global cascades on random networks,'' {\em
  Proceedings of the National Academy of Sciences}, vol.~99, no.~9,
  pp.~5766--5771, 2002.

\bibitem{lambiotte_2019}
R.~Lambiotte, M.~Rosvall, and I.~Scholtes, ``From networks to optimal
  higher-order models of complex systems,'' {\em Nature Physics}, vol.~15,
  no.~4, p.~313–320, 2019.

\bibitem{battiston2020networks}
F.~Battiston, G.~Cencetti, I.~Iacopini, V.~Latora, M.~Lucas, A.~Patania, J.-G.
  Young, and G.~Petri, ``Networks beyond pairwise interactions: Structure and
  dynamics,'' {\em Physics Reports}, vol.~874, pp.~1 -- 92, 2020.

\bibitem{giusti_clique_2015}
C.~Giusti, E.~Pastalkova, C.~Curto, and V.~Itskov, ``Clique topology reveals
  intrinsic geometric structure in neural correlations,'' {\em Proceedings of
  the National Academy of Sciences}, vol.~112, pp.~13455--13460, Nov. 2015.

\bibitem{reimann_cliques_2017}
M.~W. Reimann, M.~Nolte, M.~Scolamiero, K.~Turner, R.~Perin, G.~Chindemi,
  P.~Dłotko, R.~Levi, K.~Hess, and H.~Markram, ``Cliques of {Neurons} {Bound}
  into {Cavities} {Provide} a {Missing} {Link} between {Structure} and
  {Function},'' {\em Frontiers in Computational Neuroscience}, vol.~11, p.~48,
  June 2017.

\bibitem{santos_topological_2018}
F.~A.~N. Santos, E.~P. Raposo, M.~D. Coutinho-Filho, M.~Copelli, C.~J. Stam,
  and L.~Douw, ``Topological phase transitions in functional brain networks,''
  {\em Phys. Rev. E}, vol.~100, p.~032414, Sep 2019.

\bibitem{olfati2007consensus}
R.~Olfati-Saber, J.~A. Fax, and R.~M. Murray, ``Consensus and cooperation in
  networked multi-agent systems,'' {\em Proceedings of the IEEE}, vol.~95,
  no.~1, pp.~215--233, 2007.

\bibitem{patania_shape_2017}
A.~Patania, G.~Petri, and F.~Vaccarino, ``The shape of collaborations,'' {\em
  EPJ Data Science}, vol.~6, p.~18, Dec. 2017.

\bibitem{asch_effects_1951}
S.~E. Asch, ``Effects of group pressure on the modification and distortion of
  judgments.,'' {\em Groups, Leadership and Men}, pp.~177--190, 1951.

\bibitem{chang_co_diff_2018}
H.-C.~H. Chang and F.~Fu, ``Co-diffusion of social contagions,'' {\em New
  Journal of Physics}, vol.~20, p.~095001, Sept. 2018.

\bibitem{random_hypergraphs_2019}
G.~Ghoshal, V.~Zlatic, G.~Caldarelli, and M.~Newman, ``Random hypergraphs and
  their applications,'' {\em Physical review. E, Statistical, nonlinear, and
  soft matter physics}, vol.~79, p.~066118, 07 2009.

\bibitem{berge_hypergraphs_1989}
C.~Berge, {\em Hypergraphs: combinatorics of finite sets}.
\newblock No.~v. 45 in North-{Holland} mathematical library, Amsterdam ; New
  York: North Holland : Distributors for the U.S.A. and Canada, Elsevier
  Science Pub. Co, 1989.

\bibitem{estrada_subgraph_2005}
E.~Estrada and J.~A. Rodríguez-Velázquez, ``Subgraph centrality in complex
  networks,'' {\em Physical Review E}, vol.~71, p.~056103, May 2005.
\newblock Publisher: American Physical Society.

\bibitem{lambiotte_tda_2018}
V.~Salnikov, D.~Cassese, and R.~Lambiotte, ``Simplicial complexes and complex
  systems,'' {\em European Journal of Physics}, vol.~40, no.~1, p.~014001,
  2018.

\bibitem{schaub_random_2018}
M.~T. Schaub, A.~R. Benson, P.~Horn, G.~Lippner, and A.~Jadbabaie, ``Random
  walks on simplicial complexes and the normalized hodge 1-laplacian,'' {\em
  {SIAM} Review}, vol.~62, pp.~353--391, Jan. 2020.

\bibitem{mukherjee_random_2016}
S.~Mukherjee and J.~Steenbergen, ``Random walks on simplicial complexes and
  harmonics,'' {\em Random Structures \& Algorithms}, vol.~49, no.~2,
  pp.~379--405, 2016.

\bibitem{parzanchevski_simplicial_2017}
O.~Parzanchevski and R.~Rosenthal, ``Simplicial complexes: Spectrum, homology
  and random walks,'' {\em Random Structures {\&} Algorithms}, vol.~50,
  pp.~225--261, May 2016.

\bibitem{muhammad_control_nodate}
A.~Muhammad and M.~Egerstedt, ``Control using higher order laplacians in
  network topologies,'' in {\em Proc. of 17th International Symposium on
  Mathematical Theory of Networks and Systems, Kyoto}, pp.~1024--1038, 2006.

\bibitem{petri_simplicial_2018}
G.~Petri and A.~Barrat, ``Simplicial {Activity} {Driven} {Model},'' {\em
  Physical Review Letters}, vol.~121, p.~228301, Nov. 2018.

\bibitem{salnikov2018simplicial}
V.~Salnikov, D.~Cassese, and R.~Lambiotte, ``Simplicial complexes and complex
  systems,'' {\em European Journal of Physics}, vol.~40, no.~1, p.~014001,
  2018.

\bibitem{iacopini_2019}
I.~Iacopini, G.~Petri, A.~Barrat, and V.~Latora, ``Simplicial models of social
  contagion,'' {\em Nature Communications}, vol.~10, Jun 2019.

\bibitem{arruda2019social}
G.~F. de~Arruda, G.~Petri, and Y.~Moreno, ``Social contagion models on
  hypergraphs,'' {\em Physical Review Research}, vol.~2, Apr. 2020.

\bibitem{carletti_random_2020}
T.~Carletti, F.~Battiston, G.~Cencetti, and D.~Fanelli, ``Random walks on
  hypergraphs,'' {\em Physical Review E}, vol.~101, p.~022308, Feb. 2020.
\newblock Publisher: American Physical Society.

\bibitem{helali_hitting_2019}
A.~Helali and M.~Löwe, ``Hitting times, commute times, and cover times for
  random walks on random hypergraphs,'' {\em Statistics \& Probability
  Letters}, vol.~154, June 2019.

\bibitem{zhou_learning_2006}
D.~Zhou, J.~Huang, and B.~Schölkopf, ``Learning with hypergraphs: clustering,
  classification, and embedding,'' in {\em Proceedings of the 19th
  {International} {Conference} on {Neural} {Information} {Processing}
  {Systems}}, {NIPS}'06, (Cambridge, MA, USA), pp.~1601--1608, MIT Press, Dec.
  2006.

\bibitem{lu_high-ordered_2011}
L.~Lu and X.~Peng, ``High-{Ordered} {Random} {Walks} and {Generalized}
  {Laplacians} on {Hypergraphs},'' in {\em Algorithms and {Models} for the
  {Web} {Graph}} (A.~Frieze, P.~Horn, and P.~Prałat, eds.), Lecture {Notes} in
  {Computer} {Science}, (Berlin, Heidelberg), pp.~14--25, Springer, 2011.

\bibitem{neuhauser_multibody_2020}
L.~Neuhäuser, A.~Mellor, and R.~Lambiotte, ``Multibody interactions and
  nonlinear consensus dynamics on networked systems,'' {\em Physical Review E},
  vol.~101, no.~3, p.~032310, 2020.
\newblock Publisher: American Physical Society.

\bibitem{Srivastava2011}
V.~Srivastava, J.~Moehlis, and F.~Bullo, ``On bifurcations in nonlinear
  consensus networks,'' {\em Journal of Nonlinear Science}, vol.~21,
  pp.~875--895, Dec 2011.

\bibitem{homophily}
M.~McPherson, L.~Smith-Lovin, and J.~M. Cook, ``Birds of a feather: Homophily
  in social networks,'' {\em Annual Review of Sociology}, vol.~27, no.~1,
  pp.~415--444, 2001.

\bibitem{granovetter}
M.~Granovetter, ``Threshold models of collective behavior,'' {\em American
  Journal of Sociology}, vol.~83, no.~6, p.~1420–1443, 1978.

\bibitem{watts}
D.~J. Watts, ``A simple model of global cascades on random networks,'' {\em
  Proceedings of the National Academy of Sciences}, vol.~99, no.~9,
  p.~5766–5771, 2002.

\bibitem{asch}
S.~E. Asch, ``Group forces in the modification and distortion of judgments.,''
  {\em Social psychology.}, p.~450–501, 1951.

\bibitem{sjt}
M.~Sherif and C.~I. Hovland, {\em Social judgment: assimilation and contrast
  effects in communication and attitude change}.
\newblock Greenwood Press, 1980.

\bibitem{jager-amblard}
W.~Jager and F.~Amblard, ``Uniformity, bipolarization and pluriformity captured
  as generic stylized behavior with an agent-based simulation model of attitude
  change,'' {\em Computational and Mathematical Organization Theory}, vol.~10,
  no.~4, p.~295–303, 2005.

\bibitem{sociopatterns}
``Sociopatterns collaboration.'' \url{http://www.sociopatterns.org/}.
\newblock Accessed:2020-03-18.

\bibitem{primary_school_1}
J.~Stehlé, N.~Voirin, A.~Barrat, C.~Cattuto, L.~Isella, J.-F. Pinton,
  M.~Quaggiotto, W.~V.~D. Broeck, C.~Régis, B.~Lina, and et~al.,
  ``High-resolution measurements of face-to-face contact patterns in a primary
  school,'' {\em PLoS ONE}, vol.~6, no.~8, 2011.

\bibitem{primary_school_2}
V.~Gemmetto, A.~Barrat, and C.~Cattuto, ``Mitigation of infectious disease at
  school: targeted class closure vs school closure,'' {\em BMC Infectious
  Diseases}, vol.~14, no.~1, 2014.

\bibitem{sfhh_work}
M.~G{'e}nois and A.~Barrat, ``Can co-location be used as a proxy for
  face-to-face contacts?,'' {\em EPJ Data Science}, vol.~7, p.~11, May 2018.

\bibitem{highschool}
J.~Fournet and A.~Barrat, ``Contact patterns among high school students,'' {\em
  PLoS ONE}, vol.~9, no.~9, 2014.

\bibitem{sekara_fundamental_2016}
V.~Sekara, A.~Stopczynski, and S.~Lehmann, ``Fundamental structures of dynamic
  social networks,'' {\em Proceedings of the National Academy of Sciences},
  vol.~113, pp.~9977--9982, Sept. 2016.

\end{thebibliography}
\end{document}